\documentclass[prd,aps,floats,floatfix,superscriptaddress,preprintnumbers,
showpacs,eqsecnum,nofootinbib,twocolumn]{revtex4}
\usepackage{amsfonts,amsmath,amssymb,latexsym,array,
theorem,mathrsfs,subfigure,bm,float,color,graphicx}

\definecolor{red}{rgb}{1,0,0}
\definecolor{blue}{rgb}{0,0,1}
\definecolor{green}{rgb}{0,1,0}

\DeclareMathAlphabet{\mathpzc}{OT1}{pzc}{m}{it}

\newcommand{\ma}[1]{\mbox{$\mathcal{#1}$}}

\newcommand{\red}[1]{{{\textcolor{red}{#1}}}}
\newcommand{\blue}[1]{{{\textcolor{blue}{#1}}}}
\newcommand{\green}[1]{{{\textcolor{green}{#1}}}}

\newcommand{\calhR}[1]{\raisebox{2ex}{\tiny ({\em h})}\hspace{-0.8em}{\ma R}}

\newcommand{\lsim}{\,\mbox{
\raisebox{-1.ex}{$\stackrel{\textstyle<}{\textstyle\sim}$}}\,}

\begin{document}

\title{
Oscillating Bianchi IX Universe in Ho\v{r}ava-Lifshitz Gravity
}


\author{Yosuke {\sc Misonoh}}
\email{y"underscore"misonou"at"moegi.waseda.jp}
\address{Department of Physics, Waseda University, 
Okubo 3-4-1, Shinjuku, Tokyo 169-8555, Japan}
\author{Kei-ichi {\sc Maeda}}
\email{maeda"at"waseda.jp}
\address{Department of Physics, Waseda University, 
Okubo 3-4-1, Shinjuku, Tokyo 169-8555, Japan}
\address{Waseda Research Institute for Science and Engineering,
Okubo 3-4-1, Shinjuku, Tokyo 169-8555, Japan}
\author{Tsutomu {\sc Kobayashi}}
\email{tsutomu"at"tap.scphys.kyoto-u.ac.jp}
\address{Hakubi Center, Kyoto University, Kyoto 606-8302, Japan}
\address{Department of Physics, Kyoto University, Kyoto 606-8502, Japan}


\date{\today}

\begin{abstract}
We study a vacuum Bianchi IX universe in the context of
 Ho\v{r}ava-Lifshitz (HL) gravity. In particular,
 we focus on the classical dynamics of the universe
and analyze how anisotropy changes the history of the universe.
For small anisotropy, we find an oscillating universe as well as 
a bounce universe just as the case of 
the Friedmann-Lemaitre-Robertson-Walker (FLRW) spacetime.
However, if the initial anisotropy is large, 
we  find the universe which ends up with a big crunch after oscillations
if a cosmological constant $\Lambda$ is zero or negative.
For $\Lambda>0$, we find a variety of histories of the universe, that is 
de Sitter expanding universe after oscillations in addition to 
the  oscillating solution and the previous big crunch solution.
This fate of the universe shows sensitive dependence of
initial conditions, which is one of the typical properties of a chaotic system.
If the initial anisotropy is near the upper bound,
we find the universe starting from a big bang and ending up with a big crunch
for $\Lambda\leq 0$, while de Sitter  expanding 
universe starting from a big bang for  $\Lambda>0$. 
\end{abstract}


\pacs{04.60.-m, 98.80.Cq, 98.80.-k} 

\maketitle

\section{Introduction}

Since the advent of the big bang theory, 
the initial singularity problem is of prime importance in 
the field of cosmology.
As shown by Hawking and Penrose\cite{singularity_theorem}, 
general relativity (GR)
predicts a spacetime singularity if a certain condition is satisfied. 
Their singularity theorem concludes 
that our universe must have an initial singularity.
However, once a singularity is formed, general relativity 
becomes no longer valid. It must be replaced by more fundamental
gravity theory.
Even in the framework of an inflationary scenario which resolves many 
difficulties in the early universe based on the big bang theory,  
the initial singularity cannot be avoided.
New gravitational theory may be required to describe the beginning
of the universe.

Many researchers attempt to resolve this singularity
 problem in the context of generalization or extension of general
 relativity\cite{singularity_avoidance}. 
However, no success has been achieved yet.
Superstring theory, which is one of the most promising
candidates for unified theory of fundamental interactions,
may solve it, but so far it has not been completed yet and 
is not so far able to describe any realistic 
 strong gravitational phenomena.
Loop quantum gravity theory may resolve the problem of a big bang singularity
via loop quantum cosmology\cite{LQC}. However it is still unclear how to 
describe time evolution of quantum spacetime in loop quantum gravity 
because of the lack of ``time" variable.

Among attempts to construct a complete quantum gravitational theory, 
Ho\v{r}ava-Lifshitz (HL) gravity has been attracted much interest 
as a candidate for such a theory over the past years. 
HL gravity is characterized by its power-counting renormalizablity,
which is brought about by a Lifshitz-like anisotropic scaling as 
$t\to\ell^z t, \Vec{x}\to\ell\Vec{x}$,
with the dynamical critical exponent $z=3$ in the ultra-violet (UV) 
limit~\cite{Horava}.
In order to recover general relativity (or the Lorentz invariance)
in our world, one expects that
the constant $\lambda$ converges  to unity 
in the infrared (IR) limit in the renormalization flow.
Although it has been argued that there exist some fundamental 
problems in HL gravity
\cite{Charm,Li,BPS_1,K-A,Henn,DM,DM2,BPS_2,Papazoglou:2009fj,BPS_3,KP,addition2}, 
some extensions are proposed to remedy these 
difficulties\cite{SVW,BPS_2,general_covariance,da_Silva}.
It is intriguing issue 
whether or not HL gravity can be a complete theory of quantum gravity.

There are a number of works on cosmology in HL gravity
\cite{DM,Takahashi-Soda,scinv,cosmology6,
cosmology10,cosmology52,WangM,
cosmology13,cosmology22,cosmology27,
cosmology3,
cosmology15,
Izumi,Greenwald,non-Gaussianity,
cosmology36,Minamitsuji:2009ii,Wang:2009rw,cosmology38,
cosmology17,cosmology39,
cosmology32,Calc,Brandenberger,
Kiritsis,previous,review1,review2,review3,lambda_to_infty,
add05,
cosmology20,cosmology21,
addition1,addition3}.
As pointed out by earlier works, 
a big bang initial singularity may be avoided in the framework of
 HL cosmology due to the higher order terms in the spatial curvature
 $\mathpzc{R}_{ij}$ in the action~\cite{Brandenberger}.
In this context, many researchers have studied the dynamics of the 
Friedmann-Lemaitre-Robertson-Walker (FLRW)
universe in HL gravity
\cite{cosmology36,Minamitsuji:2009ii,Wang:2009rw,cosmology38,
cosmology17,cosmology39,
cosmology32,Calc,Brandenberger,
Kiritsis,previous,review1,review2,review3,lambda_to_infty,
addition1,addition3}. 
In isotropic and homogeneous spacetime, higher curvature terms with arbitrary 
coupling constants
mimic various types of matter with arbitrary sign of energy densities.
The $z=2$ and $z=3$ scaling terms give ``dark radiation" and
``dark stiff-matter", respectively.
Although ``dark radiation" terms in the models 
with the detailed balance condition can avoid the initial singularity,
such terms may become irrelevant to the dynamics when we include 
relativistic matter fields, which may scale as $z=3$ 
in the UV limit and behave as a stiff matter\cite{cosmology3}.

In our previous paper~\cite{previous},
we have studied the dynamics of vacuum FLRW spacetime
in generalized HL gravity model without the detailed balance condition 
and shown that ``dark stiff-matter" can avoid the initial singularity
of the universe. Even if we include relativistic matter fields,
when the contribution of ``dark stiff-matter" is dominant,
the singularity is avoided and an oscillating spacetime
 or a bounce universe is obtained.

Although we have shown a singularity avoidance in HL cosmology,
the following question may arise:
Is this singularity avoidance generic? 
Is such a non-singular spacetime stable against 
anisotropic and/or inhomogeneous perturbations?
In order to answer for these questions, 
we have to study more generic spacetime than the FLRW universe.

The initial state of the universe could be
 anisotropic and/or inhomogeneous.
Before the singularity theorem, 
some people believed that the big bang singularity 
appears because of its high symmetry and
it may be resolved if one studies anisotropic 
and/or inhomogeneous spacetime. Then they analyzed
anisotropic Bianchi-type universes and their generalization.
Although they found some interesting behaviours near the singularity 
such as chaos in Bianchi IX spacetime
\cite{chaos_IX1,chaos_IX2,chaos_IX3,chaos_IX4}, they could not succeed the
singularity avoidance. It is simply because the singularity
theorem does not allow a singularity avoidance in GR.
The situation becomes worse if we consider anisotropy 
and/or inhomogeneity.
Even in the effective gravity model derived from superstring,
which shows a singularity avoidance,  
with such a property\cite{non-singular_universe1,non-singular_universe2},
once we include  anisotropy and/or inhomogeneity, the property of such a 
singularity avoidance 
may be spoiled\cite{anisotropic_non-singular_universe}.

Therefore it is important to study 
whether or not non-singular universes in the present HL cosmology still 
exist with anisotropy and/or inhomogeneity.
In the present paper, 
we shall investigate the possibility of the singularity avoidance 
in homogeneous but anisotropic Bianchi IX universe.
Since we are interested in a singularity avoidance,
 we focus on  an oscillating universe
and analyze how anisotropy changes the history of the universe.
We will not study the chaotic behaviour in detail, which may appear near 
the big bang singularity, 
although it is one of the most popular
and important properties in the Bianchi IX spacetime
and was discussed analytically in~\cite{cosmology20,cosmology21}.
As we will show later, however,
 some property of non-integrable system, {\it i.e.}, 
sensitive dependence on initial conditions 
may be found in the fate of the universe in the present 
analysis as well.

The paper is organized as follows.
After a short overview of HL gravity,
we present the basic equations for the
 vacuum Bianchi IX universe in HL gravity
in Sec.\ref{Bianchi_IX}.
In Sec.\ref{Linear_Perturvation}, we
 study the stability of the closed FLRW universe
against small anisotropic perturbations.
In Sec.\ref{oscillating_Bianchi}
we analyze Bianchi IX universe numerically and show
 a variety of histories of the universe, depending on 
initial anisotropy.
Summary and remarks follow in Sec.\ref{summary_discussion}.
In Appendix, we also analyze a bounce universe 
with anisotropy as an another type of 
non-singular solution.

\section{Bianchi IX universe in Ho\v{r}ava-Lifshitz gravity}
\label{Bianchi_IX}
First we introduce our Lagrangian of HL gravity, by which we 
will discuss the Bianchi  IX universe.
The basic variables in HL gravity are
the lapse function, $N$, the shift vector, $N_i$,
and the spatial metric, $g_{ij}$.
These variables are subject to the action~\cite{Horava, SVW}
\begin{eqnarray}
S_{\rm HL}={1\over 2\kappa^2}
\int dt d^3x  \sqrt{g}N\left(
\mathscr{L}_K-\mathscr{V}_{\rm HL}[g_{ij}]
\right),
\label{HL_action}
\end{eqnarray}
where $\kappa^2=1/M_{\rm PL}^2$ ($M_{\rm PL}$: the Planck mass) and
the kinetic term is given by
\begin{eqnarray}
\mathscr{L}_K=\mathpzc{K}_{~ij}\mathpzc{K}^{\,ij}-\lambda \mathpzc{K}^2
\end{eqnarray}
with 
\begin{eqnarray}
&& \mathpzc{K}_{~ij}:={1\over 2N}
\left(\partial_t g_{ij}-\nabla_iN_j-\nabla_jN_i\right)
\\
&& \mathpzc{K}:=g^{ij}\mathpzc{K}_{~ij}
\end{eqnarray}
being the extrinsic curvature and its trace.
The potential term $\mathscr{V}_{\rm HL}$ will be defined shortly.
In GR we have $\lambda=1$, only for which
the kinetic term is invariant under general coordinate transformations.
In HL gravity, however, Lorentz symmetry is broken in exchange for 
renormalizability
and  the theory is invariant under
the foliation-preserving diffeomorphism transformations,
\begin{eqnarray}
t\to\bar t(t),\,~~ x^i\to\bar x^i(t,x^j).\label{f-p}
\end{eqnarray}

As implied by the symmetry ($\ref{f-p}$),
it is most natural to consider the projectable version of HL gravity,
for which the lapse function depends only on $t$: $N=N(t)$
\cite{Horava}.
Since the Hamiltonian constraint is derived from the variation
 with respect to the lapse function,
in the projectable version of the theory,
the resultant constraint equation is not imposed locally at each point 
in space,
but rather is an integration over the whole space.
In the cosmological setting, the projectability condition
results in an additional dust-like component in the Friedmann 
equation \cite{DM}.

The most generic form of the potential $\mathscr{V}_{\rm HL}$ is given 
by~\cite{SVW}
\begin{eqnarray}
\mathscr{V}_{\rm HL}&=&
2\Lambda
+g_1\mathpzc{R}
\nonumber\\&&
+\kappa^2\left(g_2\mathpzc{R}^2
+g_3\mathpzc{R}^{i}_{~j}\mathpzc{R}^{j}_{~i}\right)
+\kappa^3 g_4\epsilon^{ijk}\mathpzc{R}_{\,i\ell}\nabla_j
\mathpzc{R}^\ell_{~k}
\nonumber \\
&&+\kappa^4\Bigl(g_5\mathpzc{R}^3
+g_6\mathpzc{R}\,\mathpzc{R}^{i}_{~j}\mathpzc{R}^{j}_{~i}
+g_7\mathpzc{R}^{i}_{~j}\mathpzc{R}^{j}_{~k}\mathpzc{R}^{k}_{~i}
\nonumber\\&&
+g_8 \mathpzc{R}\Delta \mathpzc{R}
+g_9\nabla_i\mathpzc{R}_{\,jk}\nabla^i\mathpzc{R}^{jk}
\Bigr)
\,,\label{potential}
\end{eqnarray}
where $\Lambda$ is a cosmological constant,
$\mathpzc{R}^{i}_{~j}$
 and $\mathpzc{R}$ are the Ricci and scalar curvatures
of the 3-metric $g_{ij}$, respectively,
and
$g_i$'s ($i=1,..., 9$) are the dimensionless 
coupling constants.
By a suitable rescaling of time
we set $g_1=-1$.
We also adopt the unit of $\kappa^2=1$ ($M_{\rm PL}=1$)
throughout the paper.

Let us consider a Bianchi IX spacetime, which metric 
 is written as
\begin{eqnarray}
ds^2=-dt^2+{a^2 \over 4} e^{2\beta_{ij}}\omega^i\omega^j
\,,
\end{eqnarray}
where the invariant basis $\omega^i$ is given by
\begin{eqnarray}
\omega^1 &=& -\sin x^3 dx^1 + \sin x^1 \cos x^3 dx^2 \,, \nonumber \\
\omega^2 &=&  \cos x^3 dx^1 + \sin x^1 \sin x^3 dx^2 \,, \nonumber \\
\omega^3 &=&  \cos x^1 dx^2 + dx^3 \,. 
\end{eqnarray} 
A typical scale of length of the universe is given by
$a$, which reduces to the usual scale factor
in the case of the FLRW universe.
We shall call it a scale factor 
in Bianchi IX model as well.
The traceless tensor $\beta_{ij}$ measures the anisotropy of 
the universe. The spacelike sections of the Bianchi IX is
 isomorphic to a three-sphere $S^3$, and 
a closed FLRW model is a special case of the above metric
 in the isotropic limit ($\beta_{ij}\rightarrow 0$).

For a vacuum spacetime, without loss of generality, 
we can assume that $\beta_{ij}$ is diagonalized as 
\begin{eqnarray}
\beta_{ij}={\rm diag}\left(\beta_++\sqrt{3}\beta_-,
\beta_+-\sqrt{3}\beta_-,-2\beta_+\right)
\,.
\end{eqnarray}
The basic equations describing the dynamics of 
Bianchi IX spacetime 
in HL gravity are now given by the followings:
\begin{eqnarray}
&&H^2={2\over 3(3\lambda-1)}\left[3(\dot{\beta}_+^2+\dot{\beta}_-^2)
+{64\over a^6} V(a,\beta_\pm)+{8 C\over  a^3} \right]
\,,
\nonumber\\&&
\label{hamconst}
\end{eqnarray}
\begin{eqnarray}
\dot{H}+3H^2-{8\over 3(3\lambda-1)}
\left[{8\over a^5}{\partial V\over \partial a}
+{3C\over a^3}
\right]=0
\,,
\label{Friedmann2}
\end{eqnarray}
\begin{eqnarray}
\ddot{\beta}_\pm+3H\dot{\beta}_\pm
+{32\over 3a^6}
{\partial V\over \partial \beta_\pm}
=0 
\,,
\label{eq_beta}
\end{eqnarray}
where $H=\dot{a}/a$ is the Hubble expansion parameter, {\it i.e.},
the volume expansion rate is given by ${\cal K}=3H$.
The constant $C$ arises from the projectability 
condition and it
could be ``dark matter"~\cite{DM}, but
here we assume $C=0$ just for simplicity.

The potential $V$, which depends on 
$a$ as well as $\beta_\pm$, is defined by
\begin{widetext}
\begin{eqnarray}
V(a,\beta_\pm)&:=&
{a^6\over 128}
\mathscr{V}_{\rm HL}
=
V_0(\beta_\pm)+V_1(\beta_\pm)a
+V_2(\beta_\pm)a^2+V_4(\beta_\pm)a^4
+ {\Lambda \over 64}a^6
\,, \label{potential_define}
\end{eqnarray}
where
\begin{eqnarray}
V_0(\beta_\pm)
&=& e^{12\beta_{+}}\left[ -{1\over 8}( g_5 +3g_6 +g_7 -8g_9 )
\cosh (12\sqrt{3}\beta_{-} )
    +{1 \over 4}( 3g_5 +5g_6 +3g_7 -4g_9 )\cosh ( 8\sqrt{3}\beta_{-}) 
\right.
\nonumber
\\
&&
\left. - {1\over 8}( 15g_5 +13g_6 +15g_7 - 24g_9 )\cosh 
( 4\sqrt{3}\beta_{-}) +{1\over 4}( 5g_5 + 3g_6 +5g_7 -12g_9 ) \right] 
\nonumber
\\
&&
    +e^{6\beta_{+}}\left[ {1\over 4}( 3g_5 + 5g_6 + 3g_7 -4g_9 )
\cosh(10\sqrt{3}\beta_{-} ) 
    - {1\over 4}( 9g_5 +7g_6 -3g_7 +20g_9 ) \cosh ( 6\sqrt{3}\beta_{-})
 \right.
\nonumber
\\
&&
\left. + {1\over 2}( 3g_5 + g_6 -3g_7 +12g_9)\cosh ( 2\sqrt{3}\beta_{-})
\right]  
    - {1\over 8}( 15g_5 +13g_6 +15g_7 -24g_9 )\cosh (8\sqrt{3}\beta_{-} ) 
\nonumber
\\
&&
 +{1\over 2}( 3g_5 +g_6 -3g_7 +12g_9 )\cosh ( 4\sqrt{3}\beta_{-}) 
+{1\over 8}( 3g_5 +9g_6 +27g_7 -72g_9 )  
\nonumber
\\
&&
    +e^{-6\beta_{+}}\left[ {1\over 2}( 5g_5 +3g_6 +5g_7 -12g_9 )
\cosh (6\sqrt{3}\beta_{-} )
    +{1\over 2}( 3g_5 +g_6 -3g_7 +12g_9 )\cosh ( 2\sqrt{3}\beta_{-}) 
\right]  
\nonumber
\\
&&
    +e^{-12\beta_{+}}\left[-{1\over 8}( 15g_5 +13g_6 +15g_7 -24g_9) 
\cosh (4\sqrt{3}\beta_{-} )
 -{1\over 8} ( 9g_5 +7g_6 -3g_7 +20g_9 )\right]
\nonumber
\\
&&
    +{1\over 4}( 3g_5 +5g_6 +3g_7 -4g_9 )e^{-18\beta_{+}}
\cosh ( 2\sqrt{3}\beta_{-})
    -{1\over 16}( g_5 +3g_6 +g_7 -8g_9 )e^{-24\beta_{+}},
\label{potential_order0} 
\\[1em]
V_1(\beta_\pm)
&=& {g_4\over 2}
\left[ e^{10\beta_{+}}\left( \cosh (10\sqrt{3}\beta_{-}) 
-\cosh ( 6\sqrt{3}\beta_{-}) \right) 
 +e^{4\beta_{+}}\left( -\cosh (8\sqrt{3}\beta_{-} ) 
+\cosh ( 4\sqrt{3}\beta_{-}) \right) \right.
\nonumber\\
&&
\left. +{1\over 2}e^{-8\beta_{+}} -e^{-14\beta_{+}}
\cosh (2\sqrt{3}\beta_{-}) +{1\over 2}e^{-20\beta_{+}} \right],
\end{eqnarray}
\begin{eqnarray}
V_2(\beta_\pm)
&=& {1\over 16} e^{8\beta_{+}}\left[(g_2 + 3g_3)
\cosh ( 8\sqrt{3}\beta_{-})-4(g_2 + g_3)\cosh (4\sqrt{3}\beta_{-}) 
 +  3g_2 + g_3 \right] 
\nonumber\\
&& 
+{1\over 4}(g_2 + g_3)e^{2\beta_{+}}\left[ - \cosh ( 6\sqrt{3}\beta_{-} ) 
+\cosh (2\sqrt{3}\beta_{-})  \right] 
 +{1\over 8}e^{-4\beta_{+}}\left[ (3g_2 + g_3) 
\cosh (4\sqrt{3}\beta_{-}) +g_2 + g_3 \right] 
\nonumber\\
&&
-{1\over 4}(g_2 + g_3)e^{-10\beta_{+}}\cosh ( 2\sqrt{3}\beta_{-} ) 
+{1\over 32}(g_2 + 3g_3)e^{-16\beta_{+}},
\\[.5em]
V_4(\beta_\pm)
&=& -{1\over 16}
\left[ {1 \over 2}e^{4\beta_{+}}\left( 
-\cosh ( 4\sqrt{3}\beta_{-} ) +1 \right)
 +e^{-2\beta_{+}}\cosh (2\sqrt{3}\beta_{-}) 
-{1\over 4}e^{-8\beta_{+}} \right]
\label{potential_order4}
\,.
\end{eqnarray}
\end{widetext}
Although the similar potential was found in~\cite{cosmology21},
we extend it to the case without the detailed balance condition.

\section{Linear Perturbations of the FLRW universe}
\label{Linear_Perturvation}
In this section,
 we shall analyze the present system with small anisotropies
by linear perturbations of the FLRW universe. 
We discuss the stability of the oscillating FLRW universe and
present a new type of non-singular solutions.

\subsection{Oscillating Closed FLRW Universe}
First we summarize the result of the closed FLRW spacetime,
 which metric is given by
\begin{eqnarray}
ds^2=-dt^2+a^2 \left({dr^2\over 1-r^2}+r^2d\Omega^2\right)
\,.
\end{eqnarray}
We find the Friedmann equation as
\begin{eqnarray}
{1\over 2}\dot{a}^2+\mathpzc{U}(a)=0
\label{Friedmann_eq}
\,,
\end{eqnarray}
where 
\begin{eqnarray}
\mathpzc{U}(a)={1\over 3\lambda-1}\left[
1-{\Lambda\over 3}a^2
-{g_{\rm r}\over 3a^2}-{g_{\rm s}
\over 3a^4}\right]
\label{pot_FLRW}
\,.
\end{eqnarray}
The coefficients 
$g_{\rm r}$ and $g_{\rm s}$ are defined by 
\begin{eqnarray}
g_{\rm r}&:=&6(g_3+3g_2)\,,
\nonumber \\
g_{\rm s}&:=&12(9g_5+3g_6+g_7)\,.
\label{def_gr}
\end{eqnarray}

\begin{figure}[htbp]
\begin{center}
\includegraphics[width=45mm]{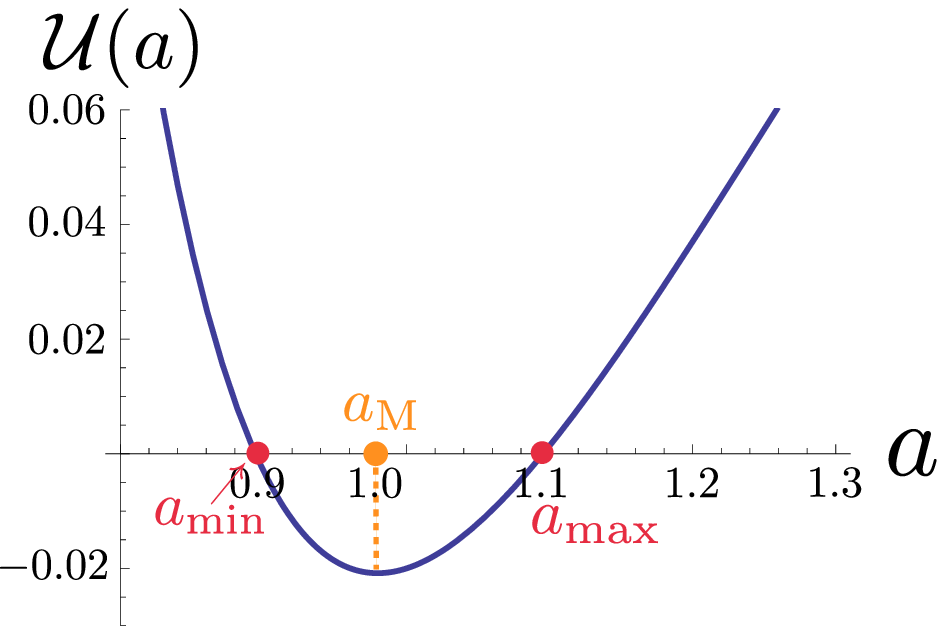} \\
(a) $\mathpzc{U}(a)$ for $\Lambda=0$ \\[1em]
\includegraphics[width=45mm]{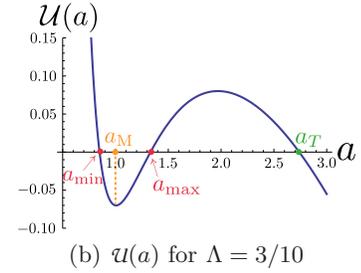} \\
(b) $\mathpzc{U}(a)$ for $\Lambda=3/10$\\[1em]
\includegraphics[width=45mm]{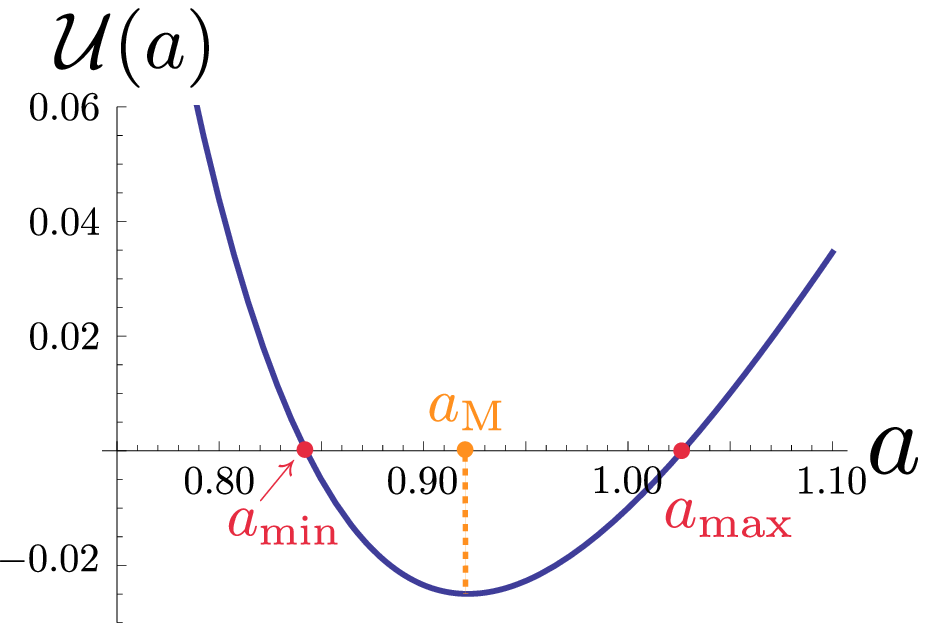} \\
(c) $\mathpzc{U}(a)$ for $\Lambda=-3/10$
\end{center}
\caption{
The potentials for the oscillating FLRW universe. 
We choose the coupling parameters as  $\lambda=1$, $g_{\rm r}=6$, and
(a) $g_{\rm s}=-72/25$,
(b) $g_{\rm s}=-72/25$,
and
(c) $g_{\rm s}=-66/25$, respectively.}
\label{pot_FLRW}
\end{figure}
The conditions for an oscillating FLRW universe to exist
were already given in~\cite{previous}, 
which are summarized as follows:\\[.5em]
(a) $\Lambda=0$
\begin{eqnarray}
g_{\rm r}>0~\,,~~~
-{g_{\rm r}^2\over 12}\leq g_{\rm s}<0
~~~~~
\,.
\label{ond_osc_FLRW}
\end{eqnarray}
(b)  $\Lambda=3/\ell^2>0$
\begin{eqnarray}
&&
\tilde g_{\rm r}>0\,,
\nonumber 
\\
&&
\tilde g_{\rm s}^{\,[1,1]\rm (-)}(\tilde g_{\rm r})
\leq 
\tilde g_{\rm s}
\left\{
\begin{array}{cc}
<& 0\\[.5em]
\leq & \tilde g_{\rm s}^{\,[1,1]\rm (+)} (\tilde g_{\rm r})
\,,
\\
\end{array}
\right.
\end{eqnarray}
(c) $\Lambda=-3/\ell^2<0$
\begin{eqnarray}
&&
\tilde g_{\rm r}>0
\,,~~~~~~~~~~~~
\nonumber \\[.5em]
&&
\tilde g_{\rm s}^{[-1,1](-)}(\tilde g_{\rm r})\leq
\tilde g_{\rm s}
<0
\,,~~~~~~~~~~~~
\end{eqnarray}
where $\tilde g_{\rm r}=g_{\rm r}/\ell^2,\,
\tilde g_{\rm s}=g_{\rm s}/\ell^4$,
and 
$\tilde g_{\rm s}^{\,[1, 1] \rm (\pm)}\left(\tilde g_{\rm r}\right)$
 is defined by
\begin{eqnarray}
\tilde g_{\rm s}^{\,[\epsilon, 1](\pm)}\left(\tilde g_{\rm r}\right)
:=
{1\over 9\epsilon^2}\left[
2-3\epsilon \tilde g_{\rm r}\pm 2 (1-\epsilon \tilde g_{\rm r})^{3/2}\right]
\label{grgs_L}
\end{eqnarray}
with $\epsilon=\pm 1$ being the sign of $\Lambda$.

We show the typical shapes of the potential $\mathpzc{U}(a)$
  in Fig. \ref{pot_FLRW}
for the coupling parameters which we use in our numerical analysis.
For an oscillating universe,
$a$ is bounded in a finite range 
as $a_{\rm min}\leq a\leq a_{\rm max}$.

\subsection{Linear Stability of the FLRW Universe}
As we mentioned in Introduction, a non-singular FLRW universe such as 
an oscillating universe should be stable against 
anisotropic perturbations. 
Otherwise such a spacetime may not be realized
in the history of the universe.
Hence, in this subsection we study stability of the FLRW universe against 
linear anisotropic perturbations.

When the anisotropic part of metric vanishes,{\it i.e.}, $\beta_\pm=0$,
Eq.~(\ref{hamconst}) reduces to the usual
Friedmann equation~(\ref{Friedmann_eq}) for a closed universe.
For the stability analysis, we expand the potential $V(a,\beta_\pm)$ around 
the FLRW universe with $\beta_\pm=0$
to second order of $\beta_\pm$, so that
\begin{eqnarray}
V_0
&\simeq&{3\over 16}
     (9g_5 + 3g_6 + g_7)
	 \nonumber\\&&
-{9\over 2} (9 g_5 - g_6 - 3 g_7 - 4 g_9)\left(\beta_+^2+\beta_-^2\right),
\\
V_1
&\simeq&
9g_4 \left(\beta_+^2+\beta_-^2\right),
\\
V_2
&\simeq&
{3\over 32}(g_3+3g_2)+{3\over 2}(g_3-3g_2)\left(\beta_+^2+\beta_-^2\right),
~~~
\\
V_4
&\simeq&
-{1\over 64}
\left[3-24\left(\beta_+^2+\beta_-^2\right)
\right]
\,.
\end{eqnarray}
The total potential $V(a,\beta_\pm)$ is thus approximated by
\begin{eqnarray}
V(a,\beta_\pm)&\simeq&
U_0(a)+U_2(a)\left(\beta_+^2+\beta_-^2\right)
\,, \label{anisotropic_expansion}
\end{eqnarray}
where
\begin{eqnarray}
U_0(a)
&=&
{3\over 16}
     (9g_5 + 3g_6 + g_7)
+{3\over 32}(g_3+3g_2)a^2
\nonumber\\&&
-{3\over 64}a^4 
+{\Lambda\over 64} a^6,
\\
U_2(a)
&=&
-{9\over 2} (9 g_5 - g_6 - 3 g_7 - 4 g_9)
+9g_4a
\nonumber\\&&
+{3\over 2}(g_3-3g_2)a^2+{3\over 8} a^4 
\label{beta_stability}
\,.
\end{eqnarray}

In order to ensure that
the FLRW universe is stable against small anisotropic perturbations,
we impose the condition $U_2(a)>0$.
The {\em sufficient} condition for stability is obtained if
all coefficients in $U_2(a)$ are positive 
because $a$ is positive, {\it i.e.}, 
\begin{eqnarray}
&&
-9 g_5 +g_6 +3 g_7 + 4 g_9\geq 0
\,, 
\nonumber
\\
&&
g_4\geq 0 \label{g_4_positive} \,, \nonumber
\\
&&
g_3-3g_2={g_{\rm r}\over 6}-6g_2\geq 0\,,
\\
\Longleftrightarrow\nonumber&&
g_2\leq {1\over 36}g_{\rm r},
\nonumber\\
&&
g_4\geq 0
\,, 
\nonumber\\
&&
g_6+g_7+g_9\geq {1\over 48}g_{\rm s}
\,.
\label{stability_beta}
\end{eqnarray}
The {\em necessary and sufficient} conditions for stability
are obtained by taking into account the dynamics of 
the background FLRW universe, {\it i.e.}, the time evolution of 
a scale factor $a$.
We will cover a wider range of the coupling parameters than the above.
To show the explicit {\em necessary and sufficient} 
conditions, just for simplicity, 
we restrict our analysis to the case with $g_4=0$, 
{\it i.e.}, the parity-conserved theory.
In this case, $U_2$ can be recast in
\begin{eqnarray}
U_2(a)&=&{3\over 8}\left[
\left(a^2-12g_2+{g_{\rm r}\over 3}\right)^2
+48(g_6+g_7+g_9)
\nonumber \right.\\
&&\left.-g_{\rm s}-\left(12g_2-{g_{\rm r}\over 3}\right)^2
\right]\,. 
\label{FLRW_stability}
\end{eqnarray}
The stability condition is such that $U_2(a)>0$ for 
$\forall a\in [a_{\rm min},a_{\rm max}]$ 
(the range in which the oscillation occurs).
When $\Lambda=0$, $a_{\rm min}$ and $a_{\rm max}$ are
given explicitly by
\begin{eqnarray}
a_{\rm min}^2&\equiv &{1\over 6}\left[
g_{\rm r}-\sqrt{g_{\rm r}^2+12g_{\rm s}}
\right]
\nonumber \\
a_{\rm max}^2&\equiv &{1\over 6}\left[
g_{\rm r}+\sqrt{g_{\rm r}^2+12g_{\rm s}}
\right]
\,.
\label{aminmax}
\end{eqnarray}
We thus find that
either of the following three conditions must be satisfied
to ensure the stability:
\begin{eqnarray}
&&
g_2\leq {1\over 72}\left(3g_{\rm r}-\sqrt{g_{\rm r}^2
+12g_{\rm s}}\right)
\,\,,
\nonumber \\
&&
g_6+g_7+g_9
\geq {1\over 72}g_{\rm s}
+{1\over 288}\left(24g_2-g_{\rm r}\right)
\nonumber \\
&&
~~~~\times
\left(g_{\rm r}-\sqrt{g_{\rm r}^2
+12g_{\rm s}}\right)
\,,
\end{eqnarray}
or
\begin{eqnarray}
&&
g_2\geq {1\over 72}\left(3g_{\rm r}+\sqrt{g_{\rm r}^2
+12g_{\rm s}}\right)
\,\,,
\nonumber \\
&&
g_6+g_7+g_9
\geq {1\over 72}g_{\rm s}
+{1\over 288}\left(24g_2-g_{\rm r}\right)
\nonumber \\
&&
~~~~\times
\left(g_{\rm r}+\sqrt{g_{\rm r}^2
+12g_{\rm s}}\right)
\,,
\end{eqnarray}
or
\begin{eqnarray}
&&
\Big{|}g_2-{1\over 24}g_{\rm r}\Big{|}
<
 {1\over 72}\sqrt{g_{\rm r}^2
+12g_{\rm s}}
\,\,,
\nonumber \\
&&
g_6+g_7+g_9
\geq {1\over 48}g_{\rm s}
+{1\over 432}\left(36g_2-g_{\rm r}\right)^2
\,.
\nonumber \\
~~
\end{eqnarray}
The above inequalities give
the necessary and sufficient conditions for 
a stable oscillating FLRW universe in the case of $\Lambda=0$ and $g_4=0$.
Since in more general cases with $\Lambda\neq 0$ and/or $g_4\neq 0$,
the necessary and sufficient conditions could
be obtained straightforwardly but
would be much more involved,
we dare not list the full conditions in the present paper.
It may be sufficient to demonstrate that the stability range of coupling 
parameters exists without any fine tuning.

\subsection{Perturbation around a static universe}
\label{perturb_static}

Next we provide a simple and illustrative example
in which even small anisotropies $\beta_\pm$ can bring 
in a possibly interesting cosmological dynamics.

Let us consider the case with 
$g_3+3g_2>0$, 
$(g_3+3g_2)^2+4(9g_5+3g_6+g_7)=0$,
and $\Lambda=0$, so that
we find a static FLRW universe with the constant 
scale factor $a=a_S:=\sqrt{g_3+3g_2}$.
We then add small anisotropic perturbations 
$\beta_\pm$ to this static background.
The basic equations governing the system are given by
\begin{eqnarray}
H^2&=&{2\over 3\lambda-1}
\left[(\dot{\beta}_+^2+\dot{\beta}_-^2)
+{64\over 3a^6} V(a,\beta_\pm)\right]
\,,~~~~~
\label{eq_fr}
\\
\ddot{\beta}_\pm
&+&{32\over 3a^6}{\partial V\over \partial \beta_\pm}
=0
\label{eq_beta}
\,,
\end{eqnarray}
where
\begin{eqnarray}
V(a,\beta_\pm)&\simeq&
U_0(a)
+U_2(a)\left(\beta_+^2+\beta_-^2\right)
\,.
\end{eqnarray}
Integrating Eqs.~(\ref{eq_fr}) and~(\ref{eq_beta}), we find
conserved anisotropic energies $E_{\beta_\pm}$ defined by
\begin{eqnarray}
E_{\beta_\pm}:={1\over 2}\left[\dot \beta_{\pm}^2+
{64\over 3a_S^6}U_2(a_S)\beta_\pm^2
\right]
\,.
\end{eqnarray}
Using the constants $E_{\beta_\pm}$
we obtain the equation for the scale factor $a$ as
\begin{eqnarray}
&&{1\over 2}\dot a^2+{4\over (3\lambda-1)a_S^2}(a-a_S)^2
={2a_S^2\over 3\lambda-1}
\left(E_{\beta_+}+E_{\beta_-}\right).
\nonumber \\
\end{eqnarray}
This equation gives an oscillating solution for $a$
with the frequency
\begin{eqnarray}
\omega_a^2:={8\over (3\lambda-1)a_S^2}.
\label{omega_a}
\end{eqnarray}
Similarly, $\beta_\pm$ also oscillate with the frequencies
\begin{eqnarray}
\omega_{\beta_\pm}^2:={64U_2(a_S)\over 3a_S^6}\,.
\label{omega_beta}
\end{eqnarray}
The ratio of two frequencies $\omega_a$ and $\omega_{\beta_\pm}$ is 
given by
\begin{eqnarray}
{\omega_{\beta_\pm}\over \omega_a}&=&\sqrt{8(3\lambda-1)U_2(a_S)
\over 3a_S^4}
\,.
\nonumber
\end{eqnarray}
Those frequencies give the typical values of the present
oscillating system.
They and their ratio are fixed only by the coupling parameters
($g_i$'s) because $a_S$ and $U_2(a_S)$ are given by them.

The above perturbative analysis around a static universe shows that
an oscillating solution newly appears in the presence of
 the small anisotropies $\beta_\pm$.
As the argument here is based on perturbations $|\beta_\pm| \ll 1$,
one may wonder whether or not there exist similar oscillating solutions 
with {\em large} anisotropies. We are going to perform numerical 
calculations to explore the anisotropic cosmological dynamics arising
 from more general setups beyond perturbations.

\section{Anisotropic oscillating universe in Ho\v{r}ava-Lifshitz gravity}
\label{oscillating_Bianchi}

In the previous section, we have shown that 
an oscillating FLRW solution in HL gravity
is stable against small anisotropies, $|\beta_\pm|\ll 1$,
for a wide range of the coupling parameters.
We now proceed to investigate the rich variety of 
the dynamics of oscillating universes
in the context of Bianchi IX spacetime, 
extending the analysis to the case with large anisotropies.
 
If $\beta_\pm$ are not small, the previous perturbative approach is 
no longer valid. 
To extend the analysis to include the case with large anisotropies, 
we employ a numerical approach and solve the governing equations 
without any perturbative expansion. With this,
we intend to uncover the rich variety of anisotropic cosmology and 
clarify the resultant fate of the universe.

We have already given the basic equations for the Bianchi IX universe 
in HL gravity.
It will be convenient to rewrite the equations as
\begin{eqnarray}
&&H^2={2\over 3(3\lambda-1)}\left[\sigma^2 
+{64\over a^6} V(a,\beta_\pm)+{8 C\over  a^3} \right] \label{basic_eq_1},
\nonumber\\&&\\
&&
\dot{H}+3H^2= {8\over 3(3\lambda-1)}
\left[{8\over a^5}{\partial V\over \partial a}
+{3C\over a^3}
\right] \label{basic_eq_2}
\,,
\\
&&
\dot{\beta}_\pm =\sigma_\pm
\label{basic_eq_3}
\\
&&
\dot{\sigma}_\pm +3H \sigma_\pm
+{32\over 3a^6}
{\partial V\over \partial \beta_\pm}
=0
\label{basic_eq_4}
\,,
\end{eqnarray}
where 
\begin{eqnarray}
\sigma^2 :={1\over 2}\sigma_{\alpha\beta}\sigma^{\alpha\beta}
= 3\left( \sigma_{+}^2 + \sigma_{-}^2 \right)
\,.
\end{eqnarray} 
$(\sigma_{\alpha\beta})={\rm diag}(\sigma_++\sqrt{3}\sigma_-,
\sigma_+-\sqrt{3}\sigma_-,-2\sigma_+)$ is the shear tensor 
of a timelike normal vector
perpendicular to the homogeneous three-space, and $\sigma$ is 
its magnitude. It may be convenient to introduce 
the dimensionless shear by
\begin{eqnarray}
\Sigma_\pm ={\sigma_\pm \over  H}~~{\rm and}~~\Sigma ={\sigma \over  H}
\,,
\end{eqnarray} 
which measure the relative anisotropies to the expansion rate $H$.
We also introduce the phase variable $\varphi$ defined by
\begin{eqnarray}
\varphi := \arctan \left({ {\sigma}_- \over  {\sigma}_+} \right)\,, 
\end{eqnarray}
which parameterizes the direction of the anisotropic expansion.

We have the five first-order evolution equations for $H$, 
$\beta_\pm$ and $\sigma_\pm$,
{\it i.e.}, Eqs.~(\ref{basic_eq_2}), (\ref{basic_eq_3})and (\ref{basic_eq_4}), 
supplemented with one constraint~(\ref{basic_eq_1}).
We have to set up the initial values for five of the six variables,
$a, H, \beta_\pm, \Sigma^2$ and $\varphi$.
The other one is fixed by the constraint equation.
Since we are interested in how the cosmological dynamics is altered
by the introduction of anisotropies,
we start with the isotropic oscillating universe
by setting $\beta_{\pm,0}$ to vanish and $a_0$ to be
a local minimum $a_M$ of ${\cal U}(a)$,
with arbitrary shear ($\Sigma_0^2$ and $\varphi_0$) at the initial moment.
So we shall give the initial data for $a_0, \beta_{\pm,0}, \Sigma_0^{~2}$ 
and  $\varphi_0$,
and then determine $H_0$ (or $\dot a_0$) by the 
constraint~(\ref{basic_eq_1}).

Without any loss of generality, we can analyze only the range 
of $0 \leq \varphi_0 \leq \pi/3$
because of the discrete symmetry modulo $\pi/3$
 of the potential $V(a,\beta_\pm)$.
Since ${\cal U}(a_M)$ is negative 
(therefore $V(a_M,\beta_\pm=0)$ is positive) for 
the oscillating FLRW universe, 
the possible range of initial shear $\Sigma_0^2$ is limited 
from  Eq.(\ref{basic_eq_2}) as
\begin{eqnarray}
0 \leq \Sigma_0^2 <\Sigma_{0{\rm (max)}}^2:= {3(3\lambda-1)\over 2}
\,.
\end{eqnarray}

As we discussed in the previous section,
the FLRW universe is stable against small anisotropic perturbations
if $U_2(a)$ defined by Eq.~(\ref{FLRW_stability}) is positive.
However one may suspect that it becomes unstable when anisotropy is large. 
For stability against large anisotropy, we have 
one natural indicator, which is the potential $V(a,\beta_\pm)$.
If the potential is unbounded from below for large 
$|\beta_\pm|$, 
we expect that if initially large $|\beta_\pm|$ 
will diverge in time and 
the universe evolves into a singularity.

From Eq.~(\ref{potential_order0}), we find 
 that the potential is bounded from below, if 
and only if one of the following conditions is satisfied:
\begin{widetext}
\begin{eqnarray}
&\mathrm{(i)}&   \ g_9 > {1 \over 8} (g_7 +3g_6 +g_5)
\,, \\
&\mathrm{(ii)}&  \ g_4\geq 0, \ 5g_7 +7g_6 +5g_5\geq 0\,,
~{\rm and}~~g_9 = {1 \over 8} (g_7 +3g_6 +g_5) 
\ \ \mathrm{with} \ \ 
 g_4^2+(5g_7 +7g_6 +5g_5)^2\neq 0\,,
\\
&\mathrm{(iii)}& \ 3g_3 +g_2 \geq 0,\ g_6\geq 0  
\,, g_4=0,\ g_7= -{1 \over 5}(7g_6 -5g_5)\,,
 \ \mathrm{and} \ g_9 = {1 \over 5}g_6 
\ \ \mathrm{with} \ \
 (3g_3 +g_2)^2+g_6^2\neq 0\,,\\
&\mathrm{(iv)}& \ g_3 \geq 0,\ g_5 \leq 0 
\,,
g_2=-3g_3,\ g_7=-g_5\,, 
 \ \mathrm{and} \ g_4=g_6=g_9=0
\,.
\end{eqnarray}
\end{widetext}
We then classify the potential $V$ into four types:
SS, US, SU, and UU, where the first S (stable) or U (unstable) 
denotes the stability against small perturbations around FLRW 
spacetime, while the second S (stable) or U (unstable) 
 corresponds to the stability against large anisotropies.
Since a cosmological constant will also change the fate of the universe,
we shall discuss eight types of cosmological models;
Models 
I-SS, I-SU, I-US, I-UU,
II-SS, II-SU, II-US, and  II-UU,
depending on the sign of $\Lambda$ 
(I for $\Lambda\leq 0$  and II 
for $\Lambda>0$)
and the potential types.
\begin{table}[b]
\begin{tabular}{|c||c|c||c|c|c|c|l|}
\hline
&&&&&&&
\\[-.5em]
Model &$\Lambda$
&$V$
&$g_3$
&$g_5$
&$g_6$
&$g_9$
&~~Figures
 \\[.5em]
\hline
\hline
&&&&&&&
\\[-.5em]
I-SS
&0
&SS
&1
&0
&$-{2\over 25}$
&${3\over 100}$
&Figs. 2-9
\\[.5em]
\hline
&&&&&&&
\\[-.5em]
I-SU
&0
&SU 
&1
&$-{11\over 225}$
&${1\over 15}$
&${1\over 100}$
&Figs. 10-12
\\[.5em]
\hline
&&&&&&&
\\[-.5em]
II-SS
&${3\over 10}$
&SS 
&1
&$-{3\over 100}$
&0
&${1\over 100}$
&Figs. 13-17
\\[.5em]
\hline 
\hline
\end{tabular}
\caption{The values of nontrivial coupling parameters $g_i$'s 
and a cosmological constant $\Lambda$, which are 
used in our numerical analysis. We also choose $\lambda=1$,
$g_2=g_4=g_7=0$. The types (SS and SU) of the potential $V$ 
are described in the text.}
\label{coupling}
\end{table}

Note that a singularity, of course,
may appear even for small anisotropies
because the Bianchi IX spacetime includes a closed FLRW model.

We have performed numerical calculations for all possible 
models and various initial data.
Now we show our numerical results for each model.
We find several types of fates of the universe
depending on the magnitude of anisotropy,
which we shall describe one by one.

In Table \ref{coupling}, we list up the values of parameters
for which we present the figures in this paper.

\subsection{Model I-SS ($\Lambda\leq 0$ and Type-SS potential)}
\label{ISS}
First we discuss Model I-SS, in which 
 $\Lambda\leq 0$ and the potential $V$ is the SS type.
Since the universe is closed, there are two fates: an eternal 
oscillation or a big crunch.
Depending on the  strength of initial anisotropy,
we find the following three types of histories of the universe.

\begin{itemize}
\item[(A)]
\underline{{\em Anisotropic oscillation}}
\,:
(small anisotropy) 
\begin{figure}[htbp]
\begin{center}
\includegraphics[width=35mm]{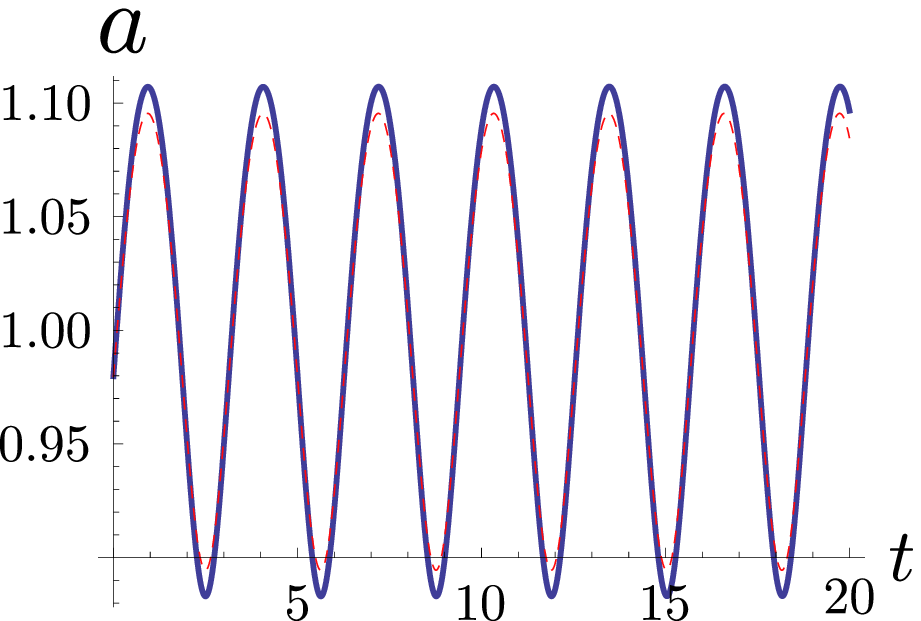} \\
(a) the time evolution of scale factor $a$ \\[1em]
\includegraphics[width=35mm]{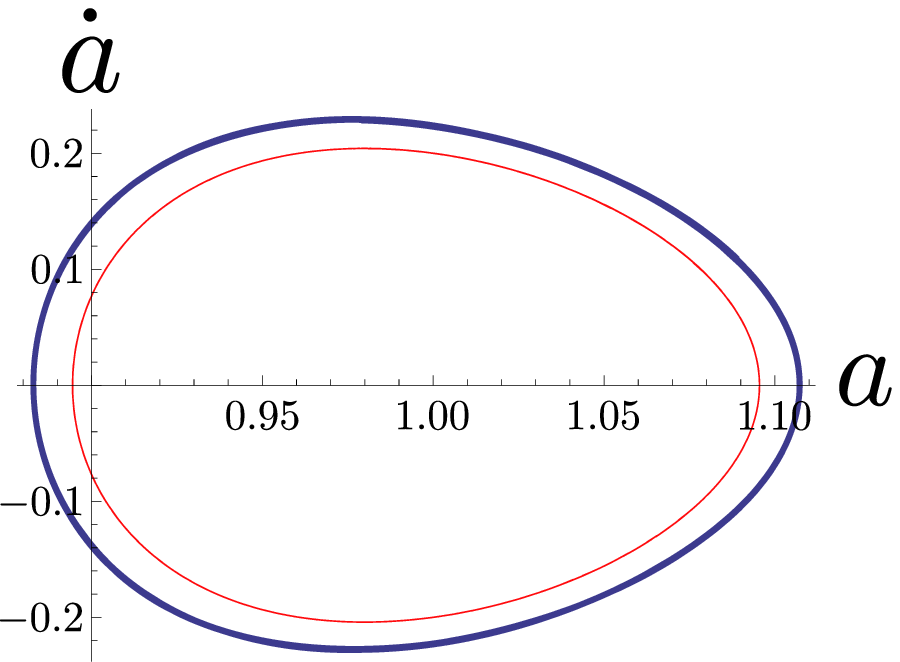} \\
(b) the phase space of $(a,\dot{a})$ 
\end{center}
\caption{The time evolution of non-singular oscillating universe.
The solid blue line and dashed red line represent Bianchi IX universe 
and FLRW universe with the same initial $a_0$, respectively. 
We show the scale factor $a$ in (a) and 
the orbit  of $a$ in
the phase space in (b), 
respectively. 
We set the initial values as 
$a_0=2\sqrt{6}/5$, $\dot{a}_0=0.2282$, 
$\Sigma_0^2=0.6000$ 
and $\varphi_0=2\pi/9$.
}
\label{cyclic_figure1}
\end{figure}
\begin{figure}[b]
\begin{center}
\includegraphics[width=45mm]{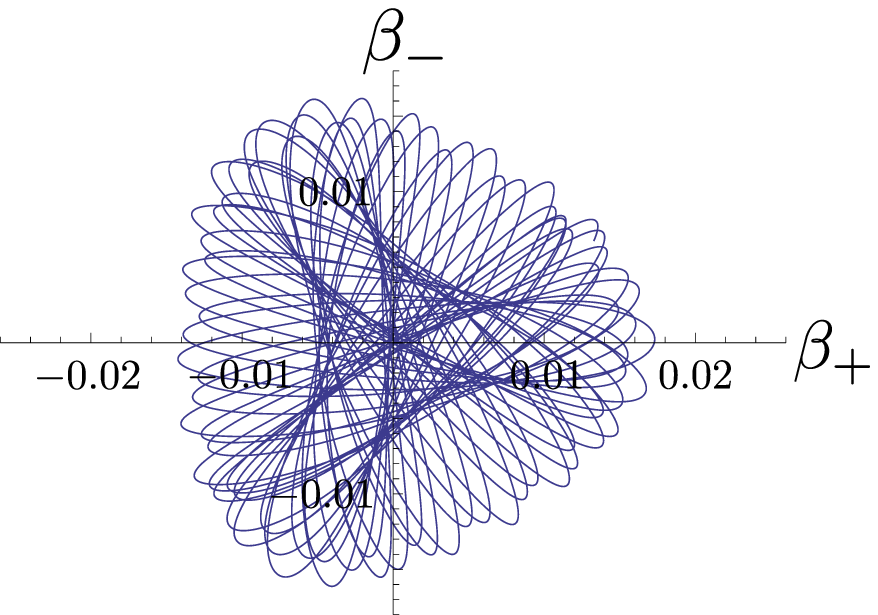}\\
(a) the orbit of ($\beta_+, \beta_-$) \\[1em]
\includegraphics[width=40mm]{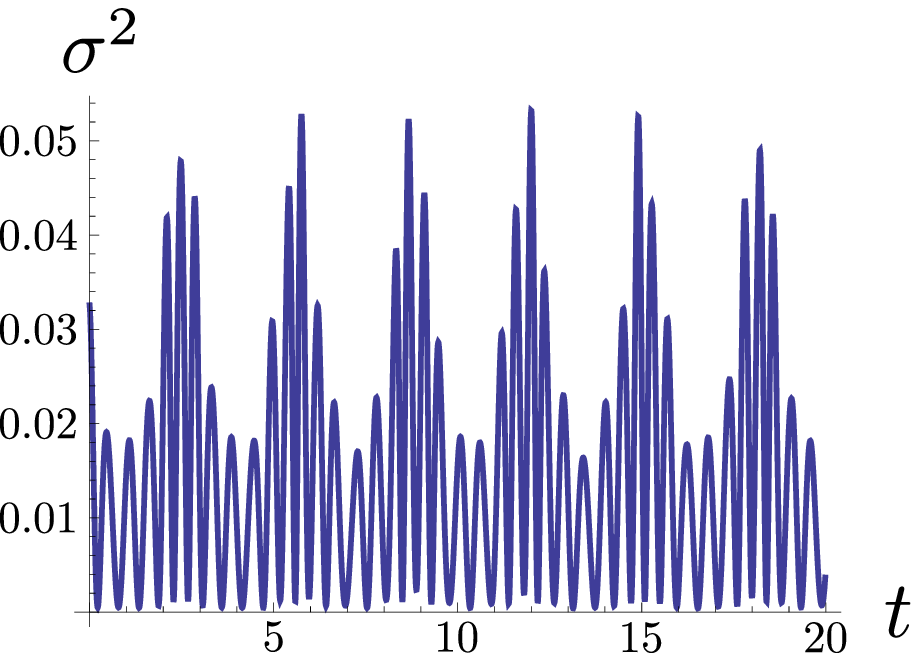}\\
(b) the evolution of the shear $\sigma^2$
\end{center}
\caption{The orbit of anisotropy 
$(\beta_+,\beta_-)$ and the time evolution of the shear $\sigma^2$
of the oscillating universe given in Fig. \ref{cyclic_figure1}.
}
\label{cyclic_figure2}
\end{figure}

Since we have an oscillating universe for the FLRW spacetime,
we find an eternally oscillating non-singular solution
 if $\Sigma_0^2$ is sufficiently small 
($\Sigma_0^2<\Sigma_{0{\rm (max)}}^2/3$).
A typical example is given in Fig.~\ref{cyclic_figure1}.
The scale factor $a$ is regularly oscillating with time just as
the FLRW solution with the same initial scale factor
 $a_0$, which is shown by  
 the dotted red curves as a reference.

This oscillating solution shows only small deviation from 
the isotropic FLRW universe.
The scale factor $a$ (and then the volume) oscillates very regularly.
Its orbit in the phase space shows an ellipse (a cross section of a torus)
(see Fig. \ref{cyclic_figure1}(b)).
The radius is slightly larger than that of the FLRW universe
because of the existence of shear (see Eq. (\ref{basic_eq_1}))

The orbit of the anisotropy $(\beta_+,\beta_-)$
is depicted in  Fig. \ref{cyclic_figure2}(a).
The anisotropic variables $\beta_\pm$ are 
trapped around the origin of $(\beta_+,\beta_-)$-space 
by the potential wall.
It looks complicated but definitely periodic.
The shear is also regularly oscillating as shown in Fig.
\ref{cyclic_figure2}(b), but 
the oscillation period is much shorter than that of
the scale factor $a$.
The oscillation amplitude of the shear $\sigma$ is  
then modulated by $a$-oscillation.
We can estimate those oscillation frequencies from 
the result in \S \ref{perturb_static}. 
Using the coupling parameters of the present model
 (see Table \ref{coupling}), we find 
$\omega_a=2$ and $\omega_{\beta_\pm}=2\sqrt{10}$ from Eqs. 
(\ref{omega_a}) and (\ref{omega_beta}),
which are almost the same as the frequencies in Figs
\ref{cyclic_figure1}(a) and \ref{cyclic_figure2}(b).

The resultant universe is regarded as an
 isotropic spacetime with small anisotropic perturbations.
The anisotropic oscillation may continue eternally.

\vskip 1cm
\item[(B)] 
\underline{\em Big crunch after oscillations}\,:
(large anisotropy)\\[1em]
If $\Sigma_0^2$ is large as
$\Sigma_{0{\rm (max)}}^2/3 < \Sigma_0^2 < \Sigma_{0{\rm (max)}}^2$,
 an initially  oscillating universe eventually 
collapses into a big crunch ($a=0$) after many oscillations
because of increase of the anisotropy.
A typical example of such a singular universe is shown 
in Fig.~\ref{oscillation_figure1}.

\begin{figure}[h]
\begin{center}
\includegraphics[width=40mm]{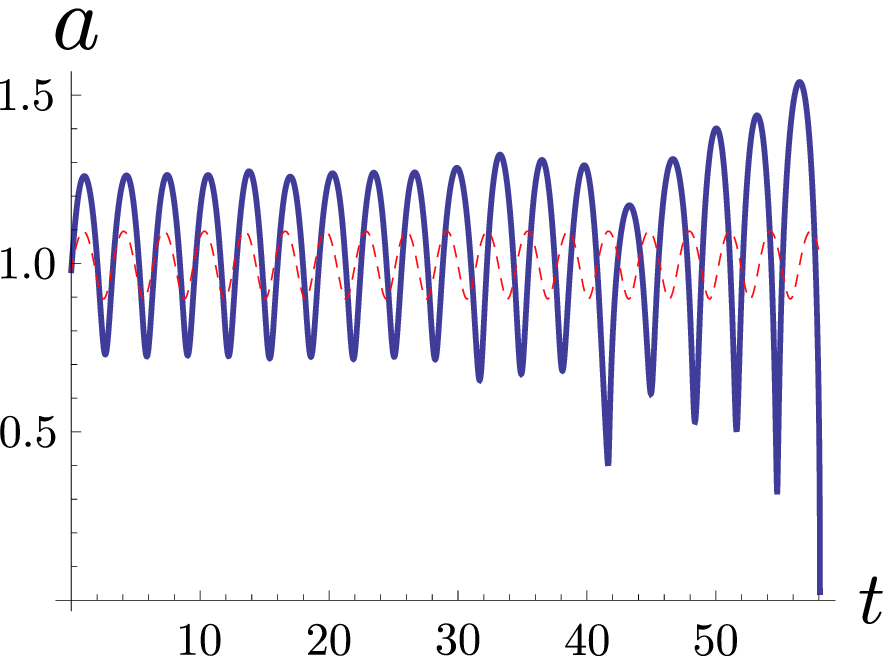} \\
(a) scale factor $a$ \\[1em]
\includegraphics[width=45mm]{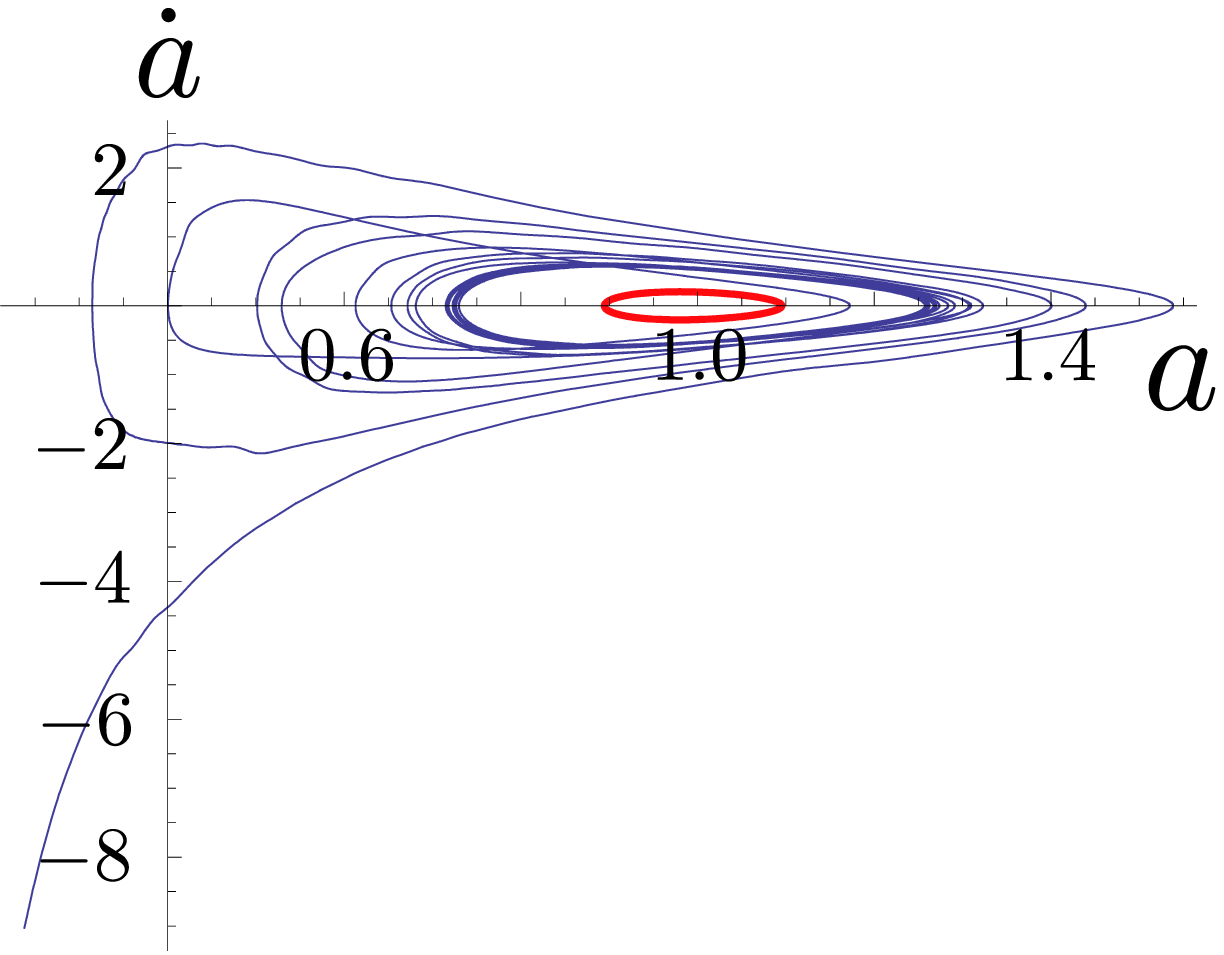} \\
(b) phase space $(a,\dot{a})$ 
\end{center}
\caption{The time evolution of the 
unstable oscillating Bianchi IX universe.
The dashed red line shows the FLRW spacetime.
We  set the initial values as $a_0=2\sqrt{6}/5$, $\dot{a}_0=0.5455$, 
$\Sigma_{0}^2=2.5800$ and $\varphi_0={2\pi/9}$.
}
\label{oscillation_figure1}
\end{figure}
The oscillation period is almost the same as 
that in Fig. \ref{cyclic_figure1}(a).
As shown in Fig.~\ref{oscillation_figure1} (b),  
the orbit of the scale factor $a$ in the phase space 
is initially almost an ellipse (a cross section
of a torus), but its ``radius" gradually increases
because of the increasing anisotropy, and 
$a$ finally evolves into a big crunch singularity 
($(a,\dot a)=(0,-\infty)$).
\begin{figure}[h]
\begin{center}
\includegraphics[width=45mm]{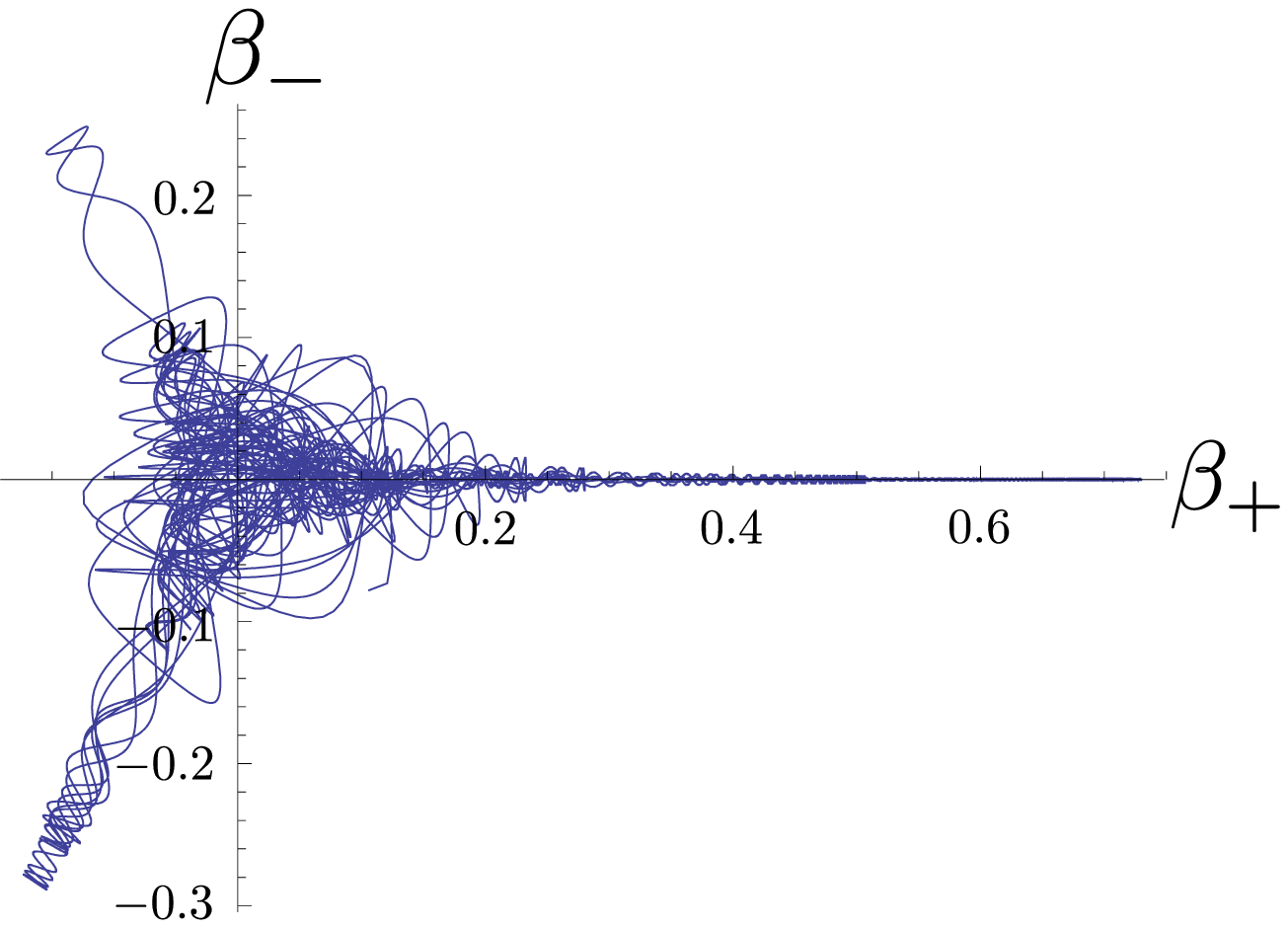} \\ 
(a) the orbit of $(\beta_+, \beta_-)$ \\[1em]
\includegraphics[width=40mm]{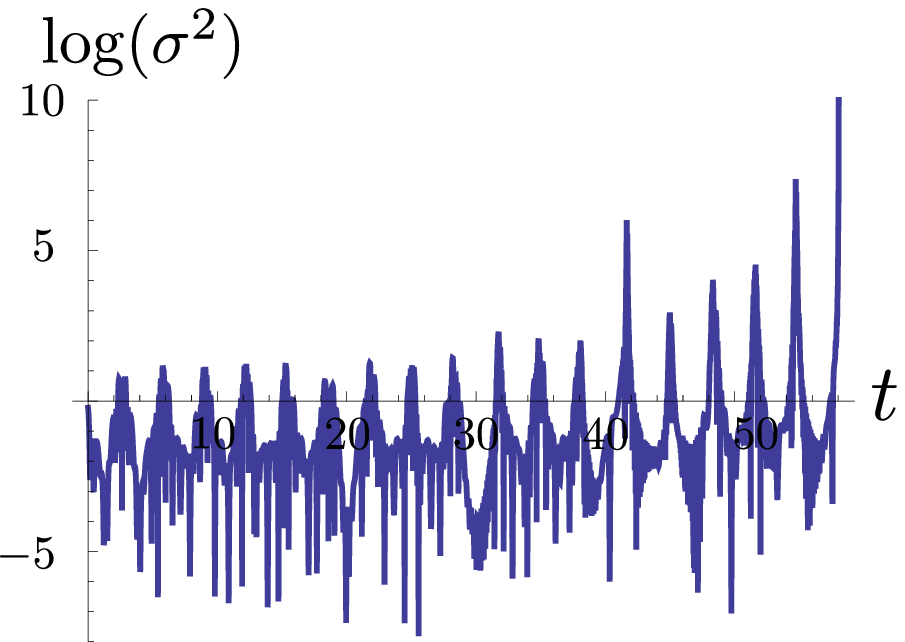} \\
(b) the evolution of shear $\sigma^2$  
\end{center}
\caption{The time evolution of the shear of the 
unstable oscillating universe given in Fig. \ref{oscillation_figure1}.}
\label{oscillation_figure2}
\end{figure}

The behaviour of anisotropy is shown 
in Fig.~\ref{oscillation_figure2}(a),
which shows that the orbit of 
 $(\beta_+,\beta_-)$ is trapped and reflected many times by the potential wall.
 The shear $\sigma$ is initially oscillating and eventually diverges 
at a big crunch  as shown in
Fig.~\ref{oscillation_figure2}(b). 
Before this divergence, we can see the increase of $\sigma^2$,
which leads the leave from the oscillating phase. 
Then the universe eventually evolves into a singularity 
with finite values of  $\beta_\pm$.
Note that the relative shear $\Sigma$ is finite at the end, which means 
that the shear is not responsible for the singularity.

In Fig. \ref{oscillation_figure_K2},
we show the curvature invariant 
\begin{eqnarray}
\mathpzc{K}_{\,ij} \mathpzc{K}^{\,ij}= 3H^2+2\sigma^2 \,,
\end{eqnarray}
 which  really diverges at a big crunch singularity.
The universe evolves into a big crunch after many oscillations.

\begin{figure}[htbp]
\begin{center}
\includegraphics[width=40mm]{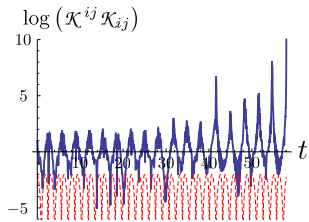}
\end{center}
\caption{The time evolution of the 
extrinsic curvature square $\mathpzc{K}^{~ij} \mathpzc{K}_{~ij}$ of the 
unstable oscillating universe given in Fig.
\ref{oscillation_figure1}. 
 For the Bianchi IX universe (solid blue line),
$\mathpzc{K}^{~ij} \mathpzc{K}_{~ij}$ oscillates in the beginning, but 
it eventually diverges at the end of the evolution, while 
it just oscillates periodically for  the  FLRW universe. 
}
\label{oscillation_figure_K2}
\end{figure}

\item[(C)]  \underline{\em From big bang to big crunch}\,:\,\\
~~(near maximally large anisotropy)\\[.5em]
~Another type of singular solution is found for 
the extremely large initial anisotropy 
($\Sigma_0^2 \sim \Sigma_{0{\rm (max)}}^2$).
In the case of the closed FLRW universe in GR,
the spacetime starts from a big bang and ends up with a big crunch. 
Bianchi IX universe in GR also evolves from zero volume 
(a big bang) to zero volume (a big crunch) through a finite maximum volume.
Hence even for the case with the oscillating FLRW universe, if we 
add a sufficiently large anisotropy, we may expect such a non-oscillating 
simple evolution.

We show one example.
As shown in Fig. \ref{sin_figure1},
the scale factor evolves from an initial finite value $a_0$
to a singularity ($a=0$), which is called a big crunch.
 If we calculate the time reversal one from the same
initial data, we will find a singularity ($a=0$), which is 
called a big bang. As a result, this universe starts from 
a big bang and ends up with a big crunch.
There is no oscillation in the evolution 
of $a$ just as the closed FLRW universe in GR.

\begin{figure}[htbp]
\begin{center}
\includegraphics[width=40mm]{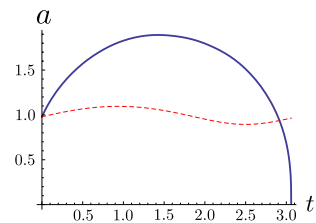}
\end{center}
\caption{The time evolution of the scale factor $a$ 
of the Bianchi IX universe with the potential given in 
Fig. \ref{potential}
 is shown by the solid blue line.
The dashed red line  represents the oscillating FLRW universe.
We set  set the initial values as 
$a_0=2\sqrt{6}/5$, $\dot{a}_0=2.0412$, 
$\Sigma_{0}^2=0.990\Sigma_{0{\rm (max)}}^2 (=2.9700)$ 
and $\varphi_0={\pi/6}$.}
\label{sin_figure1}
\end{figure}

As for the anisotropy, as shown in Fig. \ref{sin_figure2},
the orbit of $\beta_\pm$ is oscillating around the origin 
and reflecting at the potential wall until formation of 
a singularity.
 $\beta_\pm$ is finite even at a big crunch.

\begin{figure}[hbt]
\begin{center}
\includegraphics[width=50mm]{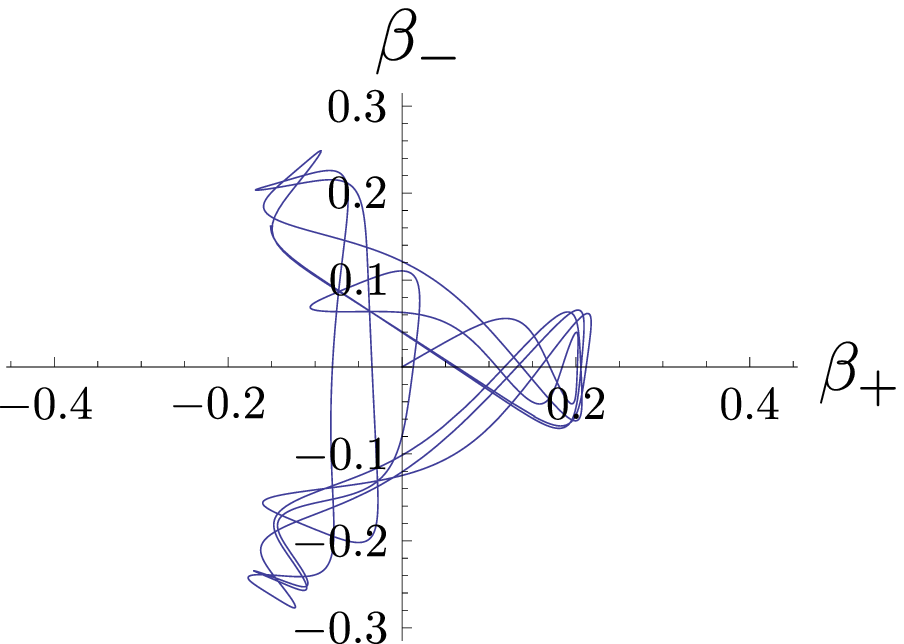} \\
(a) anisotropy ($\beta_+,\beta_-$)\\[1em]
\includegraphics[width=40mm]{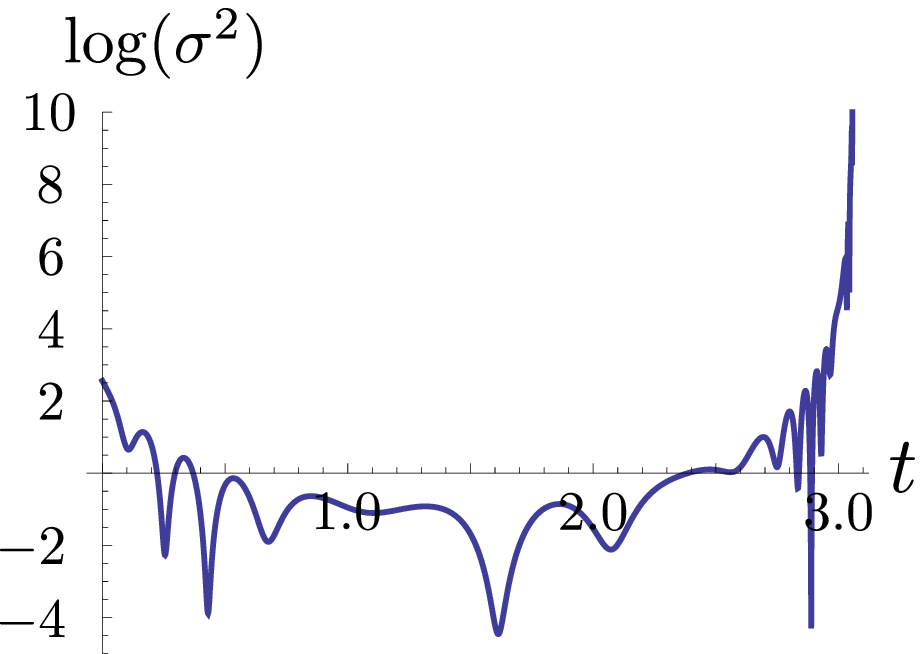}\\
(b) shear square $\sigma^2$ 
\end{center}
\caption{(a) The orbit of anisotropy ($\beta_+,\beta_-$) and 
(b) the time evolution of the shear square $\sigma^2$ 
of the universe given in Fig. \ref{sin_figure1}.
}
\label{sin_figure2}
\end{figure}

The shear $\sigma$ is also oscillating,
but the frequency is not so regular compared with 
the previous two cases (A) and (B).

The shear diverges at a big crunch ($a=0$),
which is really singular because 
 the extrinsic curvature square 
$\mathpzc{K}^{~ij} \mathpzc{K}_{~ij}$
also diverges there as
shown in Fig. \ref{sin_figureK2}.
The behaviour of $\mathpzc{K}^{~ij} \mathpzc{K}_{~ij}$
is very similar to that of the shear square.
However the relative shear $\Sigma$  does not diverge
at a big crunch, which means that the singularity 
is similar to that of the FLRW universe.
The shear does not dominate in the dynamics
(Compare it with next example (C)$'$).

\begin{figure}[htbp]
\begin{center}
\includegraphics[width=45mm]{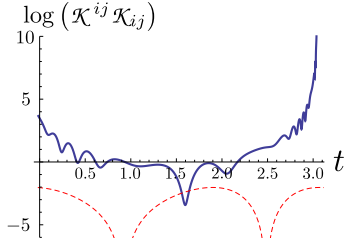}
\end{center}
\caption{The time evolution of the extrinsic curvature 
square  $\mathpzc{K}^{~ij} \mathpzc{K}_{~ij}$
 for the solution shown in Fig. \ref{sin_figure1}.
The solid blue line and dashed red line represent Bianchi IX universe 
and FLRW universe, respectively.
It diverges at the end of evolution, which is a singularity.
}
\label{sin_figureK2}
\end{figure}

In the case of a negative cosmological constant ($\Lambda<0$),
we also find the similar behaviour of the universe,
although there is a quantitative difference.

\subsection{Model I-SU ($\Lambda\leq 0$ and Type-SU potential)}
\label{ISU}
When the potential $V$ is unbounded from below,
the universe may be  unstable  against large 
anisotropic perturbations
(see Fig. \ref{ISU_potential}).
\begin{figure}[htbp]
\begin{center}
\includegraphics[width=60mm]{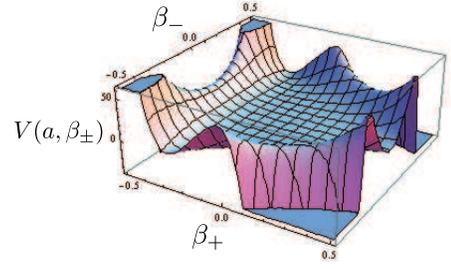} 
\end{center}
\caption{The unstable potential  $V(a,\beta_\pm)$  against large 
anisotropic perturbations is shown for  $a=1$.}
\label{ISU_potential}
\end{figure}
We also find  three types of histories of the universe
as (A), (B), and (C) in Model I-SS.
The difference from Model I-SS
appears when anisotropy gets large.
That is, the anisotropy $\beta_\pm$ diverges when a singularity appears.
We show one example with near maximally large initial anisotropy.

\item[(C)$'$] \underline{\em From big bang to big crunch}\,:\,\\
~~(near maximally  large anisotropy)\\[.5em]
In this case, just as the history (C) of Model I-SS,
the Bianchi IX universe  expands 
 from zero volume   to zero volume   
without oscillation (Fig. \ref{inf_figure1inf_figure1}).

\begin{figure}[htbp]
\begin{center}
\includegraphics[width=45mm]{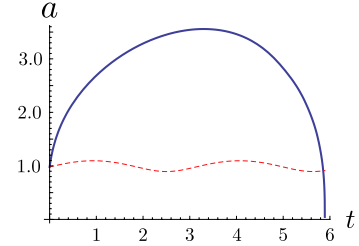}
\end{center}
\caption{The time evolution of the scale factor $a$ of the Bianchi IX universe
 is shown by the solid blue line.
The dashed red line  represents the oscillating FLRW universe.
We set  set the initial values as 
$a_0=2\sqrt{6}/5$, $\dot{a}_0=6.4549$, 
$\Sigma_{0}^2=2.9970$ 
and $\varphi_0={\pi/18}$.}
\label{inf_figure1inf_figure1}
\end{figure}
\begin{figure}[htbp]
\begin{center}
\includegraphics[width=50mm]{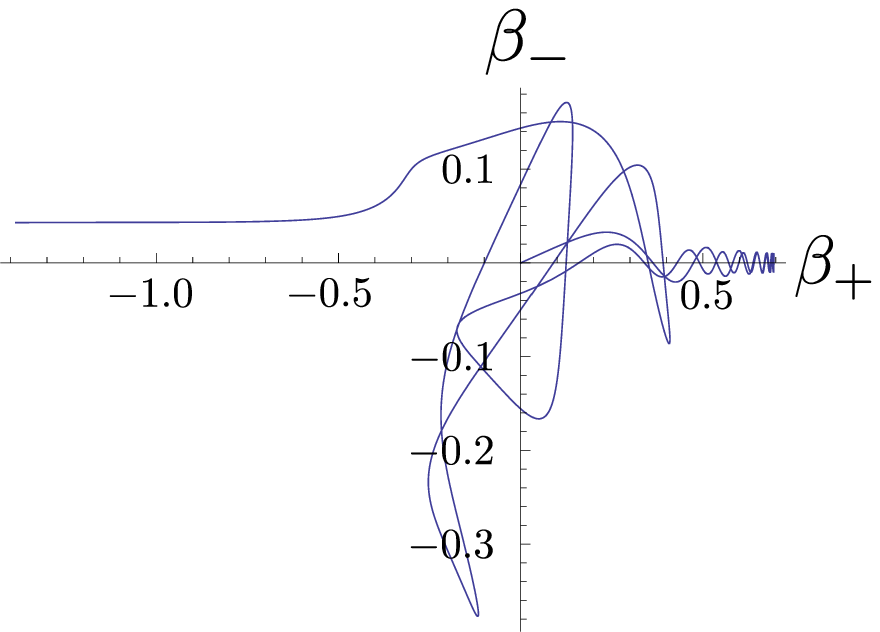} \\
(a) anisotropy ($\beta_+,\beta_-$)\\[1em]
\includegraphics[width=40mm]{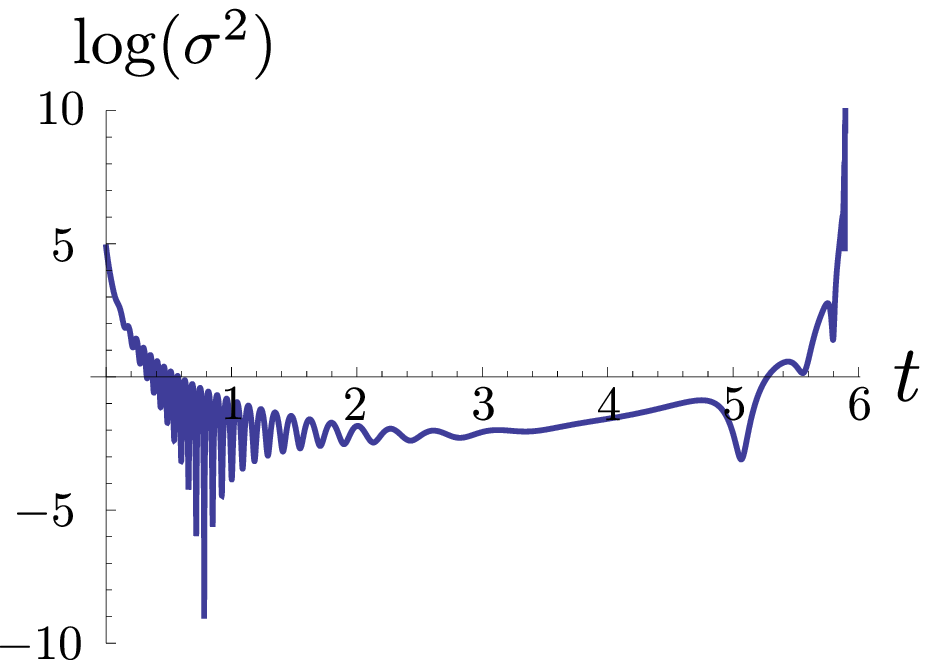}\\
(b) shear square $\sigma^2$ 
\ \\
\end{center}
\caption{(a) The orbit of anisotropy ($\beta_+,\beta_-$) and 
(b) the time evolution of the shear $\sigma^2$ 
of the universe given in Fig. \ref{inf_figure1inf_figure1}.
.}
\label{inf_figure2}
\end{figure}

The difference appears in the behaviour of  $\beta_\pm$,
which  diverges at a big crunch.
The orbit of $\beta_\pm$ initially oscillates around the origin and reflects 
on the potential wall, but it eventually evolves to infinity 
over the potential hill 
because the potential is not bounded from below
(see Fig. \ref{inf_figure2}(a)).
We also show the time evolution of the shear $\sigma^2$,
which diverges at the end of the universe.
The extrinsic curvature square  $\mathpzc{K}^{~ij} \mathpzc{K}_{~ij}$ 
also diverges there, which means it is really a singularity.

\begin{figure}[htbp]
\begin{center}
\includegraphics[width=40mm]{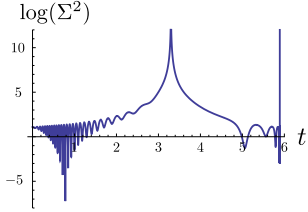}
\end{center}
\caption{
The time evolution of the relative 
shear square $\Sigma^2$ of the universe
shown in Fig. \ref{inf_figure1inf_figure1}}
\label{inf_figure3}
\end{figure}
This singularity is different from one appeared in the history (B) or (C).
To show it, we depict the time evolution of the relative shear $\Sigma$, which 
diverges at a big crunch.
It means that the shear becomes dominant at the end.
The increase of the anisotropic shear is 
responsible for the formation of a singularity.  
Note that $\Sigma$ diverges also in the middle of the
evolution but its divergence appears because of $H=0$.

At a big bang, which appears in the time reversal one, 
 we suspect that 
the shear diverges but $\beta_\pm$ is finite just as the beginning of 
Bianchi IX universe in GR.

\end{itemize}

\subsection{Model II-SS  ($\Lambda>0$ and Type-SS potential)}
\label{IISS}
Next we discuss the case of $\Lambda>0$.
In this case, we find another fate of the universe,
which is an exponentially expanding universe by a
positive cosmological constant.

If the effect of 
anisotropy is smaller than the contribution of a cosmological constant,
the universe will be isotropized.
The asymptotic equation for $a$ is given by    
\begin{eqnarray}
\ddot{a} \approx -2{\dot{a}^2 \over a} + {2 \Lambda \over 3\lambda -1}a \,,
\label{aymptotic_LambdaP}
\end{eqnarray}
if we neglect the anisotropic terms in Eq. (\ref{Friedmann2}),
 finding an exponentially expanding FLRW spacetime, {\it i.e.},
de Sitter spacetime; 
\begin{eqnarray}
a \approx e^{H_0 t}
\end{eqnarray}
with
\begin{eqnarray}
H_0 = \sqrt{{2 \Lambda \over 3(3\lambda -1)}} \,.
\end{eqnarray}
Along with the cosmic expansion
the potential is flattened to be negligible
 in the evolution equation for
the anisotropy because $a$ increases rapidly, so that 
we find 
\begin{eqnarray}
\dot{\sigma}_\pm \approx -3H_0 \sigma_\pm \,,
\end{eqnarray}
from Eq. (\ref{eq_beta}).
This implies that $\sigma_\pm\to 0$ asymptotically
 due to the Hubble friction.
Since the shear decreases to zero, the universe
becomes locally isotropic, {\it i.e.}, locally de Sitter spacetime. 
Note that this does not mean that it is
globally de Sitter spacetime because the asymptotic values 
of $\beta_\pm$ do not vanish.
However the spacetime is exponentially expanding, and  
the observable region such as a horizon scale
becomes effectively isotropic. Hence we shall 
still call this asymptotic spacetime 
 de Sitter universe.

As a result, we have three fates of the universe 
in the case with $\Lambda>0$:
oscillating universe, a big crunch, and 
de Sitter expanding universe.
We then find five 
 types of histories of the universe: 
two new types with asymptotically de Sitter universe
in addition to the previous three types (A), (B) and (C)
 discussed in \ref{ISS}.

Two new types are the similar to 
the histories (B) and (C), but different from those 
in their final states, {\it i.e.},
(D) de Sitter expansion after oscillation,
and (E) de Sitter expansion from a big bang.

For small anisotropy ($\Sigma_0^2
\lsim  \Sigma_{0{\rm (max)}}^2/3$), we find the oscillating universe (A) .
While for the large anisotropy ($\Sigma_{0{\rm (max)}}^2/3
< \Sigma_0^2<\Sigma_{0{\rm (max)}}^2$), the universe is 
either Type (B) or Type (D).
When the initial anisotropy is extremely large enough, 
{\it i.e.},  $\Sigma_0^2 \lsim \Sigma_{0{\rm (max)}}^2$, 
Type (C) or Type (E) is obtained.
We shall describe  new types (D) and (E) below.

\begin{itemize}
\item [(D)]
\underline{\em de Sitter expansion after oscillation}\,:
\\~~
(large anisotropy)

We show one example in Fig.~\ref{dS_figure1}.

\begin{figure}[htbp]
\begin{center}
\includegraphics[width=50mm]{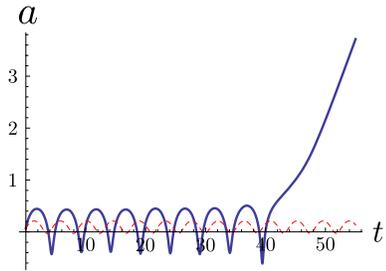}
\end{center}
\caption{The time evolution of the scale factor $a$ 
of the universe with $\Lambda>0$.
The solid blue line and dashed red line represent Bianchi IX universe
 and FLRW universe, respectively. 
We choose the coupling constants 
as $g_3=1$, $g_5=-3/100$, $g_9=1/100$,$g_2=g_4=g_6=g_7=0$, 
 $\Lambda=3/10$, and $\lambda=1$, and set the initial values as
 $a_0=1.0758$, $\dot{a}_0=0.6125$, $\Sigma_0^2=2.7000$ and 
$\varphi_0=\pi/9$.
}
\label{dS_figure1}
\end{figure}

\begin{figure}[htbp]
\begin{center}
\includegraphics[width=45mm]{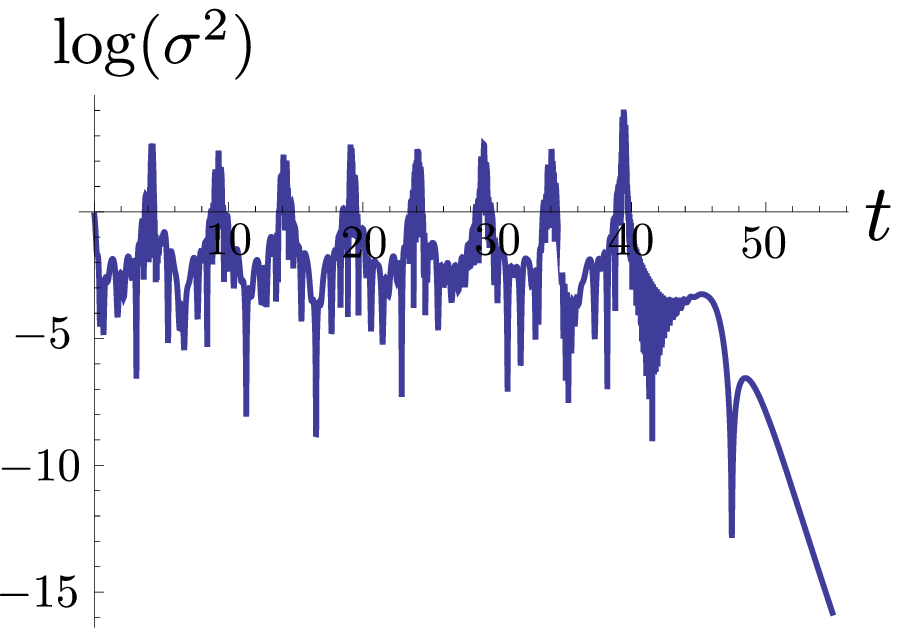} \\
(a) $\sigma^2$\\[2em]
\includegraphics[width=45mm]{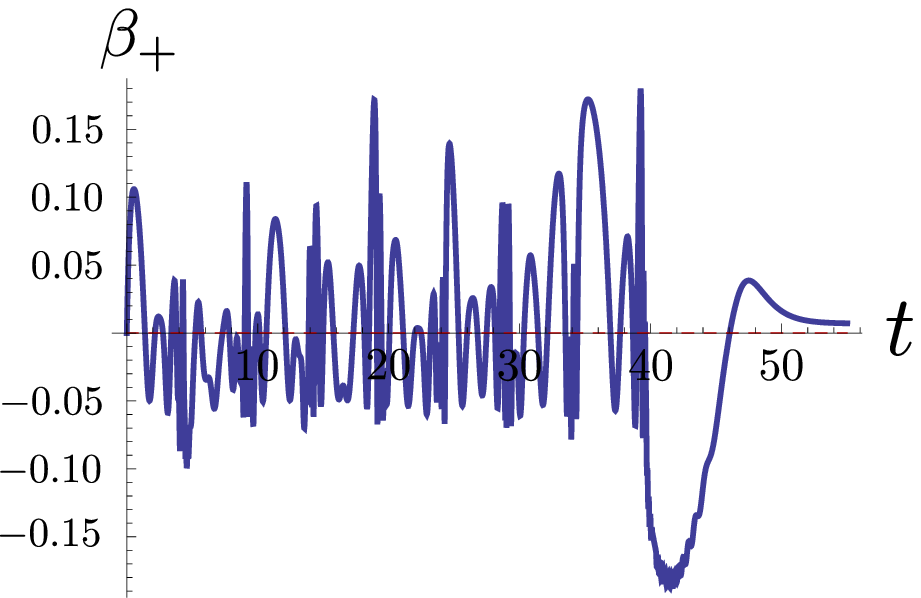}\\
(b) $\beta_+$\\[2em]
\includegraphics[width=45mm]{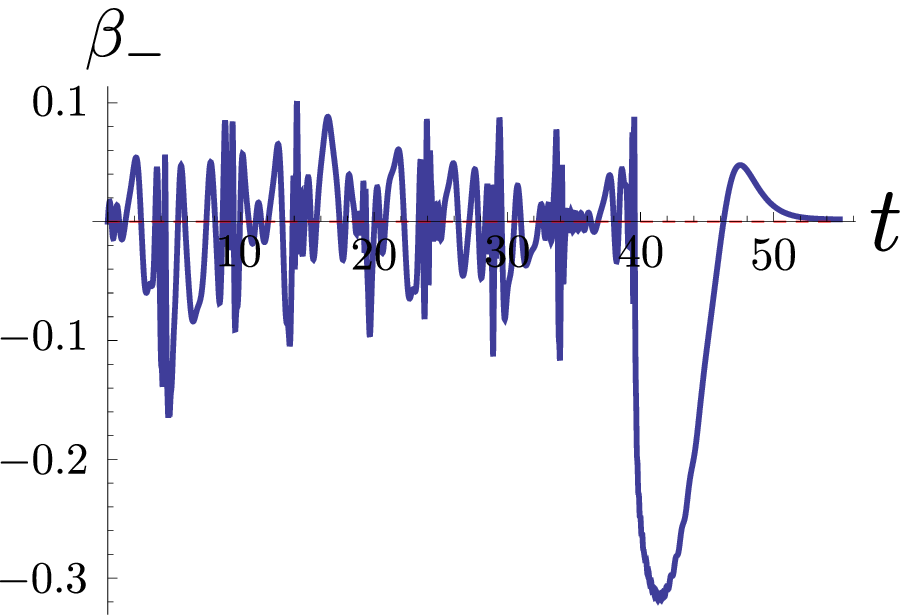} \\
(c) $\beta_-$
\end{center}
\caption{The time evolution of the shear square $\sigma^2$
and  anisotropies  $\beta_\pm$ 
of the universe shown in Fig. \ref{dS_figure1}.
The shear oscillates initially, but the universe suddenly 
leaves the oscillating phase to de Sitter expanding phase, which 
 drops the shear to zero rapidly.
The initially oscillating $\beta_\pm$ 
finally settles to finite values after small bump.}
\label{dS_figure2}
\end{figure}
If  $\Sigma_0^2$ is as large as $\Sigma_{0{\rm (max)}}^2/3<
 \Sigma_0^2< \Sigma_{0{\rm (max)}}^2$,
 an initially  oscillating universe eventually 
 evolves into an exponentially expanding 
de Sitter universe 
because of a cosmological constant.

The initially oscillating universe leaves the oscillation phase
when the anisotropy increases beyond some critical value.
We also show the evolution of the shear $\sigma$ in 
Fig.\ref{dS_figure2}.
We find that it is oscillating regularly 
for two-third of the whole period,
but eventually increases.
Then the universe leaves the oscillating phase and 
evolves into de Sitter phase.
Because of rapid expansion of the universe,
the shear vanishes soon \cite{footnote1}.

We also show the time evolution of the anisotropy
($\beta_\pm$) in Fig. \ref{dS_figure2}(b),(c).
The initially oscillating anisotropy 
increases as a burst and then decreases
to a small finite constant.

How the universe choose its fate ((B) or (D)) is as follows:
If the spacetime is expanding when it leaves from the oscillating phase,
it evolves into de Sitter phase (D), while if it is contracting,
it collapses to a big crunch (B).

\vskip .5cm

\item [(E)]
\underline{\em de Sitter expansion from big bang}\,:
\\~~
(near maximally large anisotropy)
\\[.5em]
If $\Sigma_0^2$ is close to the maximum value ($\Sigma_{0{\rm (max)}}^2=3$)
and the universe is initially expanding ($\dot{a}_0>0$),
the spacetime evolves into de Sitter phase without oscillation.
The large anisotropy makes a jump from the oscillating phase 
to de Sitter phase in the beginning.
The anisotropy drops quite rapidly because of the
exponential expansion 
as shown in Fig. \ref{dS2_figure2}.

\begin{figure}[ht]
\begin{center}
\includegraphics[width=45mm]{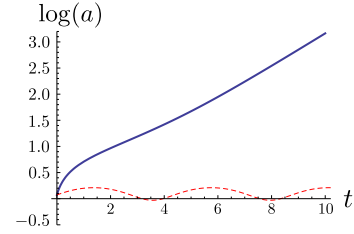}
\end{center}
\caption{The time evolution of the scale factor $a$ 
of the universe with $\Lambda>0$ and Type-SU potential.
The universe starts from a big bang and evolves into
de Sitter spacetime.
The  dashed red line represents the oscillating 
FLRW universe 
as reference. 
}
\label{dS2_figure1}
\end{figure}

On the other hand, if the universe is initially contracting 
($\dot{a}_0<0$), the spacetime is classified into Type (C),
{\it i.e.}, from a big bang, which appears 
in the time reversal one,  to a big crunch without oscillation.

\begin{figure}[htbp]
\begin{center}
\includegraphics[width=50mm]{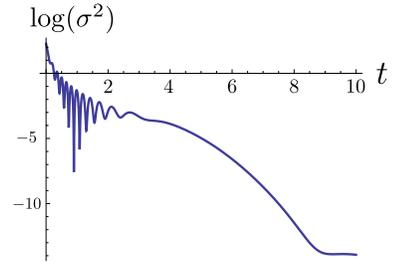} 
\end{center}
\caption{The time evolution of the shear $\sigma^2$
of the universe shown in Fig. \ref{dS2_figure1}.
Initially oscillating shears drops to zero after de Sitter expansion
starts. }
\label{dS2_figure2}
\end{figure}

\subsection{Model II-SU  ($\Lambda>0$ and Type-SU potential)}
\label{IV}
In this case, we also find the similar histories of the universe
to Types (A), (B), (C), (D) and (E),
depending on initial anisotropies.
The differences between Models II-SS and II-SU
are qualitatively the same as those 
 between Models I-SS and I-SU.
Only one difference from Model I is that there exists de Sitter phase 
as the fate of the universe because of a positive cosmological constant.

\subsection{Models with the unstable potential against 
small perturbations around FLRW spacetime}
\label{U}

In the previous four subsections, we discuss the cosmological models
with the stable potential against small perturbations around FLRW
spacetime. 
When the potential is unstable against small perturbations around FLRW
spacetime, we also find qualitatively similar results.
The main difference is that oscillations around the FLRW spacetime
never happen. Even if the universe starts from near FLRW spacetime,
it evolves into spacetime with large anisotropy
because the FLRW spacetime is unstable.
As a result, in the case of Type-UU potential,
the universe collapses to a singularity for Model I-UU.
No oscillating phase is found.
If $\Lambda>0$, {\it i.e.}, for Model II-UU,
some universe collapses to a singularity without oscillations,
and the other one evolves into 
de Sitter expanding universe,
depending on initial conditions.

\end{itemize}

\begin{widetext}
\subsection{Dependence of anisotropy on the date of the universe}
\label{summary}
In Table~\ref{oscillation_result}, we summarize the fate of the universe.
We assume the coupling parameters by which there exists
an oscillating FLRW universe.
For Models I-SS and II-SS,
we find an oscillating FLRW universe with anisotropy
in the case of small initial anisotropy. 
When we increase the strength of anisotropy,
 the spacetime leaves the initially oscillating phase 
and eventually evolves into a singularity 
or de Sitter spacetime.
If the initial anisotropy is near the maximum value, 
the oscillating phase disappears 
and a simply expanding and contracting universe is found just as 
a closed universe in GR for $\Lambda\leq 0$.
When $\Lambda>0$, an initially expanding universe 
evolves into de Sitter spacetime, while an initially
contracting universe evolves into a big crunch.

\begin{table*}[hb]
\begin{tabular}{|c|c|c|lclcl|}\toprule
Model &cosmological &
potential  &
&&
~~~~~~~~~~~~~~~~~~$\Sigma_0^2 $   && 
\\
\cline{4-8}
 &constant&
$V$ &
~~~small~~~ &$\rhd$&
~~~~~~~~~~~~~~~~~~large~~~ &$\rhd$&
~~~near maximally large~~~
\\ \hline \hline
&&&&&&&\\[-.7em]
I-SS &&
 SS &~~~(A) \green{OSC}~~~ &$\rhd$&
~~~(B) \green{OSC}$\rightarrow$\red{SING$1$}
&$\rhd$&
~~~(C) \red{SING$1$} 
 \\[.3em] 
I-US&
$\Lambda\leq 0$ 
&
US
&
~~~(A) \green{OSC}~~~&
$\rhd$& 
~~~(B) \green{OSC}$\rightarrow$\red{SING$1$}
&$\rhd$ 
&
~~~(C) \red{SING$1$} 
 \\[.3em]
I-SU &
&
SU
&
~~~(A) \green{OSC}~~~&&
~~~~~~~~~~~~~~~~~~$\rhd$ &&
~~~(C)$'$ \red{SING$2$} 
 \\[.3em]
I-UU &
&
UU
&
&&
~~~(C)$'$ \red{SING$2$} &&
 \\ \hline \hline
&&&&&&&\\[-.7em]
II-SS &
&
SS&
~~~(A) \green{OSC}~~~&
$\rhd$& 
~~~(B)$\cdot$(D) \green{OSC}$\rightarrow$\blue{deS}/\red{SING$1$}
&$\rhd$&
~~~(E) \blue{deS} [(C) \red{SING$1$}]~~~ 
 \\[.3em]
II-US
&
$\Lambda>0$ &
US&
~~~(A)  \green{OSC}~~~&
$\rhd$& 
~~~(B)$\cdot$(D) \green{OSC}$\rightarrow$\blue{deS}/\red{SING$1$}~~~
&$\rhd$ &
~~~(E) \blue{deS} [(C) \red{SING$1$}]~~~ 
\\[.3em]
II-SU&
&
SU
&
~~~(A) \green{OSC}~~~&&
~~~~~~~~~~~~~~~~~~$\rhd$&&
~~~(E)$'$ \blue{deS} [(C)$'$ \red{SING$2$}]~~~ 
 \\[.3em]
II-UU &
&
UU
&
&&
~~~(E)$'$ \blue{deS} [(C)$'$ \red{SING$2$}]~~~&&
 \\ \hline \hline
\end{tabular}
\caption{Classification of Bianchi IX cosmological models by 
a cosmological constant $\Lambda$ and types of the potential.
OSC, SING$1$, SING$2$ and de S represent
  the anisotropic oscillation, two types of big crunch singularities 
(one with  finite anisotropy and the other with infinite anisotropy), 
and de Sitter spacetime, respectively. \blue{deS}/\red{SING$1$}
means that the spacetime evolves either de Sitter phase or 
a big crunch with finite anisotropy.   \blue{deS} [\red{SING$1$} or 
\red{SING$2$}] denotes that the fate is either de Sitter universe 
if the universe is initially expanding or a big crunch if contracting.}
\label{oscillation_result}
\end{table*}

\end{widetext}

For Models I-US and II-US, the histories of the universes 
are similar to those in Models I-SS and II-US, respectively, although
deviation from isotropy becomes large even for initially small anisotropy.

In the cases of Models I-SU and II-SU, 
unless an initial anisotropy is small, an 
initially expanding universe turns to contract and collapses into a big crunch
for $\Lambda \leq 0$, while it evolves into de Sitter spacetime for 
$\Lambda>0$. This singularity at a big crunch is different from that in
Models I-SS,-US, and II-SS, US. 
The anisotropy $\beta_\pm$ for the present model 
diverges, while that for the other cases is finite even at a singularity.

Models I-UU and II-UU are not so interesting. 
There is no oscillating phase. 
If $\Lambda\leq 0$, the spacetime is simply a spacetime evolving
from a big bang to a big crunch.
For $\Lambda>0$, the initially expanding universe evolves into de Sitter 
spacetime, while the contracting one collapses into a singularity.

\begin{figure}[hb]
\begin{center}
\includegraphics[width=60mm]{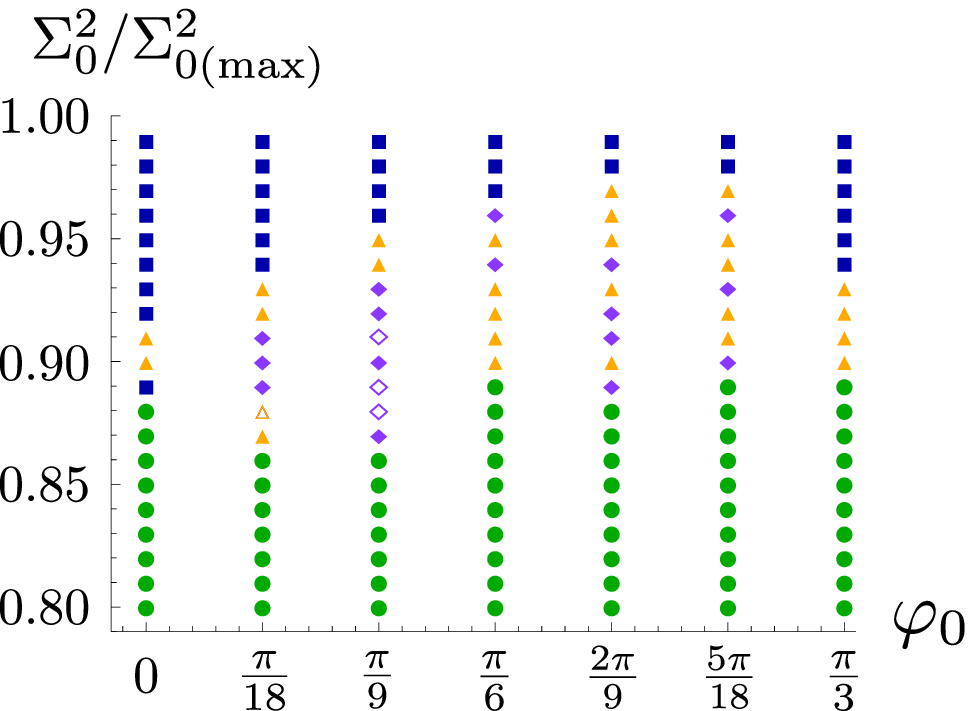} \\[.5em]
(a)  $\dot{a}_0>0$
\\
\includegraphics[width=60mm]{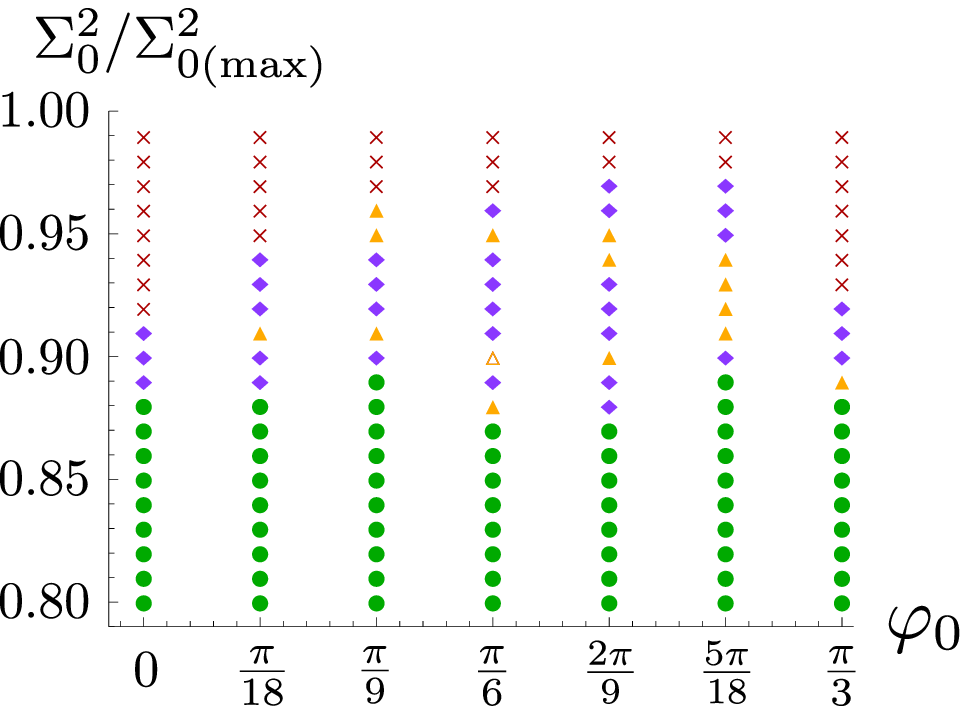}\\[.5em]
(b)  $\dot{a}_0<0$
\end{center}
\caption{The fate of Type II-SS Bianchi IX universe in terms of 
initial anisotropy $\Sigma_0^2$ and $\varphi_0$.
We judge the fate of the universe at $t=100 t_{\rm PL}$ ($t_{\rm PL}$:
the Planck time).
The histories (A), (B), (C), (D) and (E) are represented by 
a filled green circle, yellow  triangle, red cross, 
purple diamond and blue square, respectively.
The empty yellow triangle and purple 
diamond are classified to the histories
 (B) and (D), respectively, but
those oscillating periods are longer than $100t_{\mathrm{PL}}$.
We have set $a_0=1.0758$. $\dot a_0$ is fixed by the constraint
equation [(a) $\dot a_0>0$ and (b) $\dot a_0<0$].
}
\label{dS_map}
\end{figure}

One may wonder whether the initial anisotropy 
classifies the fate of the universe.  
Are there any critical values of the initial anisotropies for 
their transitions in Table \ref{oscillation_result} ?
We understand naively that 
such a transition occurs as the anisotropy increases.

We find there exists a critical value for the transition (B) to
(C) or (D) to (E), which is about $(0.94-0.98)\times \Sigma_{0{\rm (max)}}^2$.
For the transition from (A) to (B) (or (D)), it is not so clear 
whether there exists a critical value of $\Sigma_{0}^2$ or not.
If  $\Sigma_{0}^2$ is sufficiently small, we find the history (A), while
when it is large, we find the history (B) or (D). 
However, because we analyze the system numerically, 
we are not sure whether the model with small anisotropy 
oscillates forever or will turn to collapse long after.
If the latter case is true, the case with small anisotropy
is classified into the history (B) or (D).
So there may not be exactly the history (A) 
except for the exact FLRW spacetime.

More interesting fact is found in the history (B)$\cdot$(D) for $\Lambda>0$.
In order to study how the fate of the universe depends on the initial data,
we solve the basic equations in Model II-SS assuming various initial values 
of anisotropy ($\Sigma_0^2$ and $\varphi_0$). 
We summarize the results in Fig.~\ref{dS_map}.
As we see from Fig.~\ref{dS_map}, the spacetime with $\dot a_0>0$
starting from 
near maximum anisotropy evolves into de Sitter universe
(the history (E)),
while it collapses into a singularity if $\dot a_0<0$
(the history (C)).
For small $\Sigma_0^2$, we find an oscillating universe
(the history (A)).
If  $\Sigma_0^2$ is between the above two cases, however,
the fate of the universe is not so simple.
The history of such a universe is classified either (B)
OSC$\rightarrow$SING1 or (D) OSC$\rightarrow$deS.
However such a history does not shift monotonically 
from (B) to (D) as the initial anisotropy increases.
How the universe choose its fate ((B) or (D)) as follows:
If the spacetime is expanding when it leaves from the oscillating phase,
it evolves into de Sitter phase (D), while if it is contracting,
it collapses to a big crunch (B).
As a result, the fate of the universe 
is sensitively dependent on initial conditions.
If one changes the initial conditions, the fate changes drastically.
It is because the present system is non-integrable.
Such a property is found in a dynamical system with chaos.
Since our model is Bianchi IX, which shows chaotic behaviour near 
singularity in GR\cite{chaos_IX2,chaos_IX3,chaos_IX4}, 
we understand why we find such a complicated
 basin structure of the fate in Fig.~\ref{dS_map}, 
which can be fractal\cite{chaos_IX4}.

\section{Summary and remarks}
\label{summary_discussion}
We have explored a singularity avoidance in a vacuum Bianchi IX universe 
in HL gravity.
We have studied an oscillating cosmological solution with anisotropy.
In the case of small anisotropy ($|\beta_\pm| \ll 1$),
 we find an analytical solution and show the stability condition of
the FLRW universe against anisotropic perturbations. 
We have also solved the basic equations numerically 
and discussed the possible history of the universe.
We classify our models into eight types 
I-SS,-SU,-US,-UU, and II-SS, -SU,-US,-UU,
depending on the sign of a cosmological constant and 
the types of the potential $V$.
We find five types of the histories of the universe:
(A) an oscillating universe with anisotropy, (B) 
a big crunch after oscillations, (C) from a big bang to a big crunch, 
(D) de Sitter expansion after oscillations and (E) from a big bang to
de Sitter expansion, as summarized in Table \ref{oscillation_result}.

The stable oscillating universe (A) is found if initial anisotropy is small 
in the case that 
the coupling parameters ($g_i\,'$s)
 satisfy the stability condition.
When initial anisotropy is large,  
the oscillating universe evolves 
 into a singular big crunch (B) for  $\Lambda \leq 0$.
In the case of $\Lambda>0$, if the initial anisotropy is large but not 
close to the maximum value, we find two histories (B) and (D).
Which history is realized does not depend monotonically on 
the initial shear $\Sigma_0^2$, but the present system shows
sensitive dependence of initial conditions just as 
one of the typical properties of chaos.
The anisotropic bounce universe is also 
obtained for the model satisfied the stability condition 
if initial anisotropy is small.

Since we adopt the unit of $M_{\rm PL} =1$, 
the oscillation period and oscillation amplitude 
are  the Planck scale, unless the coupling
constants are unnaturally large.
Hence in order to obtain a macroscopic universe,
we need a positive cosmological constant $\Lambda>0$,
which provides us a de Sitter expanding phase.

In a more realistic situation, this cosmological constant 
should be replaced by a potential $V_\phi$ of an inflaton scalar field $\phi$.
Reheating after inflation may give an initial state of a macroscopic 
big bang universe.

When we include a scalar field, however, 
we have to take into account 
modification of a scalar field action in the UV limit
similar to HL gravity action.
The  action $S_\phi$ may be given by 
\begin{eqnarray}
S_{\phi} := 
\int dt d^3x N \sqrt{g}\left(
{\dot{\phi}^2 \over 2N^2} - \phi \mathcal{O} \phi -V_{\phi}(\phi)
\right)\,,
\label{scalar_action}
\end{eqnarray}
where 
\begin{eqnarray}
\mathcal{O} := {C_3 \over M^4} \Delta^3 +{C_2 \over M^2} \Delta^2 + C_1 
\Delta \,
\end{eqnarray}
with $M$ being a typical  mass scale and
 $C_i\,'$s ($i=1-3$) being dimensionless constants. 
For this action with HL gravity action (\ref{HL_action}), 
the basic equations for a Bianchi IX cosmological model
are given by 
\begin{eqnarray}
&&H^2={2\over 3(3\lambda-1)}\left[3(\dot{\beta}_+^2+\dot{\beta}_-^2)
+{64\over a^6} V(a,\beta_\pm)  \right. \nonumber \\
&&\left. \ \ \ \ \ \ \ \ +{8 C\over  a^3}+{1\over 2} \dot{\phi}^2 
+V_\phi(\phi) \right]
\,,~~~~
\\
&&
\dot{H}+3H^2 = {8\over 3(3\lambda-1)}
\left[{8\over a^5}{\partial V\over \partial a}
+{3C\over a^3}
+{3\over4}V_\phi(\phi)
\right]
\,,~~~~~~
\\
&&
\ddot{\beta}_\pm+3H\dot{\beta}_\pm
+{32\over 3a^6}
{\partial V\over \partial \beta_\pm}
=0 
\,,
\\
&&
\ddot{\phi}+3H\dot{\phi}
+{\partial V_\phi \over \partial \phi}
=0 
\,.
\end{eqnarray}
Although the action (\ref{scalar_action}) contains 
higher spatial derivatives, there exists no difference from the conventional
canonical kinetic term of a scalar field for a homogeneous spacetime.
We expect a usual inflationary scenario once de Sitter exponential 
expansion starts.
There exists reheating after  slow-roll inflation, finding 
a big bang universe.

If we have an oscillating phase
 before inflation, we may expect one interesting effect, which is
modification of primordial perturbations.
We should stress that a classical transition from an oscillating phase to 
an inflationary stage never happens in the FLRW model
(See our discussion in \cite{previous} for quantum transition). 
So anisotropy may be important in the pre-inflationary 
oscillating phase. 
As a result, we may find large non-Gaussian density perturbations.
To confirm our scenario, 
we should explore the dynamics of a scalar field in the pre-inflationary era 
because $V_\phi$ in an inflationary model may be fluctuated 
by the oscillating scale  factor in the pre-inflationary era while $\Lambda$
 is a constant in our present analysis.

Another interesting possibility is anisotropic inflation,
which is discussed in the model with higher curvature 
terms\cite{anisotropic_inflation1} 
or  with a vector field\cite{anisotropic_inflation2}.
It may leave distinguishable imprints on the primordial perturbations.

\acknowledgments
TK would like to thank RESCEU, the University of Tokyo,
where a large part of this work was completed.
This work was supported in part by JSPS Grant-in-Aid for Research 
Activity Start-up No. 22840011 (TK), 
by JSPS  Grant-in-Aids for Scientific Research Fund No.22540291 (KM)
and by JSPS under the Japan-Russia Research Cooperative
Program (KM).


\appendix
\section{Anisotropic bounce universe in Ho\v{r}ava-Lifshitz gravity}
\label{bounce_Bianchi}

In the text, we have considered oscillating universes
as a possible way to avoid an initial singularity.
In this appendix, we study another way to singularity avoidance,
{\it i.e.}, a bounce solution~\cite{previous}.
For a closed FLRW universe, a bounce solution exists
only only in the case of $\Lambda>0$.
As a result, a bounce solution with anisotropy
 is also found only for the case of $\Lambda>0$.

\begin{figure}[hbtp]
\begin{center}
\includegraphics[width=45mm]{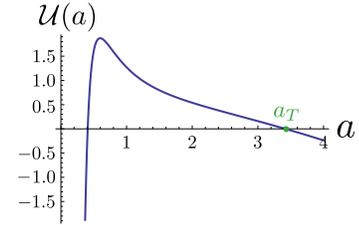}\\
(a) Type A\\[.5em]
\includegraphics[width=45mm]{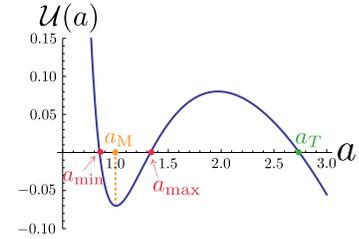}\\
(b) Type B
\end{center}
\caption{Typical shapes of potential $\mathcal{U}(a)$.
(a) Type A: We choose  the coupling constants  as 
$g_{\rm r}=-6$, $g_{\rm s}=27/25$ and 
$\Lambda=3/10$. The bounce point in the isotopic FLRW model is given by 
$a_T = 3.4183$.  (b) Type B: We choose 
the coupling constants as $g_{\rm r}=6$,
 $g_{\rm s}=-81/25$ and 
$\Lambda=3/10$. The bounce point, the maximum and minimum oscillating 
radii in the isotopic FLRW model are given by $a_T = 2.7452$, 
$a_\mathrm{max}=1.2337$, and $a_{\mathrm{min}}=0.9702$, respectively.
}
\label{bounce_potential}
\end{figure}

We classify bounce solutions into two classes 
according to the shape of the potential ${\cal U}(a)$.
The first class is such that $\mathcal{U}(0)=-\infty$, which we call
Type A, and the other is such that $\mathcal{U}(0)=+\infty$, which we call 
Type B.
We show two typical potentials in Fig.~\ref{bounce_potential}.
For Type-A potential, in addition to a bounce solution, we find a 
FLRW spacetime 
starting from a big bang and collapsing to a big crunch, while for 
Type-B potential, we have an oscillating FLRW solution as well as 
a bounce spacetime.
This difference gives rise to different fates of anisotropic 
Bianchi IX universe as we will show later.

When initial anisotropy is small, it is easy to find
anisotropic bounce solutions because 
anisotropy  can be treated as
perturbations around the isotropic closed FLRW universe.
In order for this type of stable 
solutions to exist, 
the condition $U_2(a)>0$ ($\forall a>a_T$) must be satisfied.
The sufficient condition for stability 
is the same as Eq. (\ref{stability_beta}). 

If the stability condition  $U_2(a)>0$ 
is not satisfied, we do not usually obtain 
bounce solutions. Only for extremely small initial shear
($\Sigma_0^2 \lsim 10^{-3}$),
a bounce solution can be found 
even if  the condition is not satisfied.
It is because the the bounce occurs before 
the unstable mode grows enough. 

In Table \ref{coupling2}, we list up the values of parameters
for which we present the figures and tables here.
\begin{table}[h]
\begin{tabular}{|c||c|c|c|c|c||c|}
\hline
&&&&&&
\\[-.5em]
$\mathcal{U}$&$\Lambda$
&$g_2$
&$g_3$
&$g_5$
&$g_9$
&Figures \& Tables
 \\[.5em]
\hline
\hline
&&&&&&
\\[-.5em]
A
&${3\over 10}$
&$-{1\over 6}$
&$-{1\over 2}$
&${1\over 100}$
&${3\over 100}$
&Figs. 18(a), 19, Table IV
\\[.5em]
\hline
&&&&&&
\\[-.5em]
B
&${3\over 10}$
&$0$
&$1$
&$-{3\over 100}$
&${1\over 100}$
&Figs. 18(b), 20, Table V
\\[.5em]
\hline 
\hline
\end{tabular}
\caption{The values of nontrivial coupling parameters $g_i$'s 
and a cosmological constant $\Lambda$, which are 
used in our numerical analysis. We also choose $\lambda=1$,
$g_2=g_4=g_7=0$. The types (A and B) of the potential $\mathcal{U}$ 
are described in the text.}
\label{coupling2}
\end{table}

\subsection{Type-A bounce universe}
\begin{figure}[htbp]
\begin{center}
\includegraphics[width=55mm]{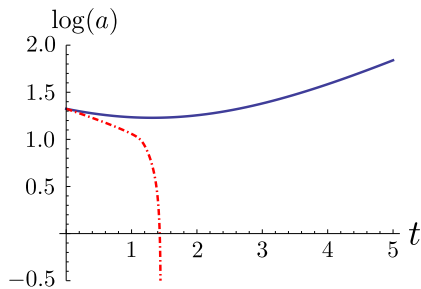}
\end{center}
\caption{The time evolutions of the scale factor $a$ of 
the universe with Type-A potential.
The solid blue, and dashed red lines 
denote de Sitter universe via a bounce, and a collapsing universe,
 respectively.
We have set $a_0=3.7601$,and $\varphi_0={5\pi/18}$,
 and  the shear square  $\Sigma_0^2=0.6900$ (solid blue) 
and  $\Sigma_0^2=1.9500$ (dash-dotted red).
 $\dot{a}_0$ is determined by the constraint equations as
 $\dot{a}_0=-0.5968$ (solid blue) and   $\dot{a}_0=-0.8852$ 
(dash-dotted red).
}
\label{bounceA_figure}
\end{figure}

First we show the results for Type-A potential in Fig.~\ref{bounceA_figure},
where we find  two typical 
evolutions of the universe. 

We assume the coupling parameters which guarantee
the existence of the FLRW bounce universe.
We then include anisotropy  and study how anisotropy
changes the fate of the universe.
We find the following results:
\begin{itemize}
\item
If the shear is small, we still have a regular bounce solution
(the solid blue curve).
\item
When the shear becomes large,
the initially contacting universe just collapses to a singularity
(the dash-dotted red line). No singularity avoidance is obtained.
\end{itemize}
Hence, there seems to exist a critical value of 
 $\Sigma_{0{\rm (crit)}}^2$, beyond which the spacetime 
collapses to a singularity.
The critical values $\Sigma_{0(\mathrm{crit})}^2$ for 
Type-A potential
are listed in Table~\ref{bounceA_critical},
which is strongly dependent on
 the initial scale factor $a_0$. 
However, the corresponding values of $\sigma_{0\rm{(crit)}}^2$
 are not so much different.
Hence we may conclude that 
the critical value is determined by the 
absolute value of the shear but not by the relative value to
the Hubble parameter. 

\begin{table}[h]
\begin{tabular}{c|c|cccccccc}
\toprule 
&  
&
&
&
&
$\varphi_0$&
&
&
& \\[.5em]
\cline{3-10}
$a_0$&shear 
&
&
&
&
&
&
&
&\\[-.5em]
& 
&
$0$ &
${\pi\over 18}$ &
${\pi\over 9}$ &
${\pi\over 6}$ &
${2\pi\over9}$ &
${5\pi\over 18}$ &
${\pi\over 3}$ \\[.5em] \hline
&&&&&&&&\\[-.7em]
& 
$\Sigma_{0{\rm (crit)}}^2$ &
$1.539$ &
$0.963$ &
$0.861$ &
$0.786$ &
$0.738$ &
$0.708$ &
$0.699$ 
 \\
$1.1 a_T$&&&&&&&&\\[-.7em]
& 
$\sigma_{0{\rm (crit)}}^2$ &
$0.426$ &
$0.285$ &
$0.264$ &
$0.246$ &
$0.237$ &
$0.231$ &
$0.228$ 
 \\
&&&&&&&&\\[-.7em]
 \hline
&&&&&&&&\\[-.7em]
& 
$\Sigma_{0{\rm (crit)}}^2$&
$0.225$ &
$0.114$ &
$0.102$ &
$0.093$ &
$0.087$ &
$0.081$ &
$0.081$
 \\
 $1.5 a_T$&&&&&&&&\\[-.7em]
& 
$\sigma_{0{\rm (crit)}}^2$&
$0.207$ &
$0.144$ &
$0.135$ &
$0.129$ &
$0.126$ &
$0.120$ &
$0.120$
 \\
&&&&&&&&\\[-.7em]
\hline 
\hline
\end{tabular}
\caption{The critical relative shear square $\Sigma_{0{\rm (crit)}}^2$ 
and the corresponding absolute shear square $\sigma_{0{\rm (crit)}}^2$
for Type-B bounce universe.
for Type-A bounce universe.
If the initial shear exceeds this critical value, 
the universe evolves into singularity.
}
\label{bounceA_critical}
\end{table}

\subsection{Type-B bounce universe}
In this case, we find the following three evolutionary histories
of the universe:\\
\begin{itemize}
\item A simple bounce solution just as Type A\\
This is possible if  the deviation from
 the ``background'' isotropic universe is sufficiently small.
\item A big crunch solution  just as Type A\\
The universe collapses into a singularity if 
$\Sigma_0^2$ is larger than some critical value $\Sigma_{0({\rm crit})}^2$.
\item A bounce solution after some oscillations\\
We also find an oscillating phase before a bounce, for which $\Sigma_0^2$ 
is very close to $\Sigma_{0({\rm crit})}^2$ initially. 
This type of solution requires a fine-tuning to some degree,
and hence in this sense the solutions are not generic.
\end{itemize}

\begin{figure}[t]
\begin{center}
\includegraphics[width=55mm]{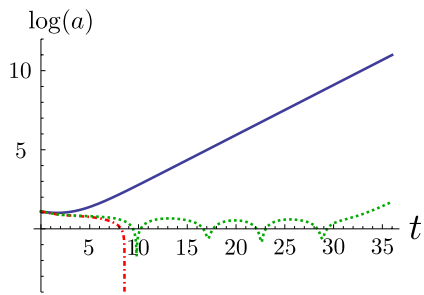}
\end{center}
\caption{The time evolutions of the scale factor $a$ of 
 the universe with Type-B potential.
The solid blue, dash-dotted red, and dotted green lines 
denote de Sitter universe via a bounce, a collapsing universe, and 
de Sitter universe after several oscillations, respectively.
We have set $a_0=3.0197$, $\varphi_0=\pi/9$, 
 and  the shear square  $\Sigma_0^2=0.9000$ (solid blue),
  $\Sigma_0^2=1.1700$ (dash-dotted red),
and $\Sigma_0^2=1.1409$ (dotted green).
 $\dot{a}_0$ is determined by the constraint equations as
$\dot{a}_0=-0.4109$ (solid blue),
$\dot{a}_0=-0.4402$ (dash-dotted red) and
$\dot{a}_0=-0.4367$ (dotted green).
}
\label{bounceB_figure}
\end{figure}

The typical evolution of Type-B universe is presented in 
Fig.~\ref{bounceB_figure}.
The critical values $\Sigma_{0{\rm (crit)}}^2$ for 
Type-B bounce universes
are listed in Table~\ref{bounceB_critical}.
As Type-A bounce solution, 
the critical value of $\sigma_{0{\rm (crit)}}^2$ does 
not strongly depend on  the initial scale factor $a_0$,
but $\Sigma_{0{\rm (crit)}}^2$ does.

Near the critical anisotropy, we suspect that 
which fate of the universe
is realized may depend sensitively on initial conditions.

\newpage
\begin{table}[h]
\begin{tabular}{c|c|cccccccc}\toprule 
&
&
&
&
&
$\varphi_0$&
&
&
& \\[.5em]
\cline{3-10}
$a_0$&
shear&
&
&
&
&
&
&
&\\[-.5em]
& 
&
$0$ &
${\pi\over 18}$ &
${\pi\over 9}$ &
${\pi\over 6}$ &
${2\pi\over9}$ &
${5\pi\over 18}$ &
${\pi\over 3}$ \\[.5em] \hline
&&&&&&&&\\[-.7em]
& 
$\Sigma_{0{\rm (crit)}}^2$ &
$1.254$ &
$1.188$ &
$1.143$ &
$1.086$ &
$1.065$ &
$1.062$ &
$1.137$ 
 \\
$1.1 a_T$&&&&&&&&\\[-.7em]
& 
$\sigma_{0{\rm (crit)}}^2$ &
$0.288$ &
$0.276$ &
$0.267$ &
$0.255$ &
$0.252$ &
$0.252$ &
$0.264$ 
 \\
&&&&&&&&\\[-.7em]
 \hline
&&&&&&&&\\[-.7em]
& 
$\Sigma_{0{\rm (crit)}}^2$&
$0.153$ &
$0.135$ &
$0.120$ &
$0.114$ &
$0.105$ &
$0.105$ &
$0.123$
 \\
$1.5 a_T$&&&&&&&&\\[-.7em]
& 
$\sigma_{0{\rm (crit)}}^2$&
$0.150$ &
$0.141$ &
$0.132$ &
$0.129$ &
$0.123$ &
$0.123$ &
$0.135$
\\
&&&&&&&&\\[-.7em]
\hline 
\hline
\end{tabular}

\caption{The critical relative shear square $\Sigma_{0{\rm (crit)}}^2$ 
and the corresponding absolute shear square $\sigma_{0{\rm (crit)}}^2$
for Type-B bounce universe.
If the initial shear exceeds  this critical value, 
the universe evolves into singularity.
}
\label{bounceB_critical}
\end{table}



\end{document}